\documentclass[
    aps,
    pre,
    twocolumn,
    superscriptaddress,
    preprintnumbers,
    amsmath,
    amssymb,
]{revtex4-2}

\usepackage[group-separator={,}]{siunitx}
\usepackage{amssymb}
\usepackage{tikz}
\usepackage{lipsum}
\usepackage{mathtools}
\usetikzlibrary{positioning}
\usetikzlibrary{angles,quotes}
\usepackage{bm}
\usepackage{empheq}

\usepackage{comment}

\newcommand{\de}{\mathrm{d}}
\DeclareMathSymbol{\shortminus}{\mathbin}{AMSa}{"39}

\begin{document}

\title{Stochastic Multistability of Clonallike States in the Eigen Model: a Fidelity Catastrophe}

\author{Emanuele Crosato}
\thanks{These two authors contributed equally}
\affiliation{School of Physics, UNSW, Sydney, NSW 2052, Australia}
\affiliation{EMBL Australia Node in Single Molecule Science, School of Biomedical Sciences, UNSW, Sydney, NSW 2052, Australia}
\affiliation{Living Systems Institute and Department of Mathematics and Statistics, University of Exeter, Exeter EX4 4QD, United Kingdom}

\author{Richard E. Spinney}
\thanks{These two authors contributed equally}
\affiliation{School of Physics, UNSW, Sydney, NSW 2052, Australia}
\affiliation{EMBL Australia Node in Single Molecule Science, School of Biomedical Sciences, UNSW, Sydney, NSW 2052, Australia}

\author{Richard G. Morris}
\email{r.g.morris@unsw.edu.au}
\affiliation{School of Physics, UNSW, Sydney, NSW 2052, Australia}
\affiliation{EMBL Australia Node in Single Molecule Science, School of Biomedical Sciences, UNSW, Sydney, NSW 2052, Australia}
\affiliation{ARC Centre of Excellence for the Mathematical Analysis of Cellular Systems, UNSW Node, Sydney, NSW 2052, Australia.}

\begin{abstract}
    The Eigen model is a prototypical toy model of evolution that is synonymous with the so-called error catastrophe: when mutation rates are sufficiently high, the genetic variant with the largest replication rate does not occupy the largest fraction of the total population because it acts as a source for the other variants.
    Here we show that, in the stochastic version of the Eigen model, there is also a \emph{fidelity} catastrophe.
    This arises due to the state-dependence of fluctuations and occurs when rates of mutation fall beneath a certain threshold, which we calculate.
    The result is a type of noise-induced multistability whereupon the system stochastically switches between short-lived regimes of effectively clonal behavior by different genetic variants.
    Most notably, when the number of possible variants--- typically $\sim4^L$, with $L\gg 1$ the length of the genome--- is significantly larger than the population size, there is only a vanishingly small interval of mutation rates for which the Eigen model is neither in the fidelity- nor error-catastrophe regimes, seemingly subverting traditional expectations for evolutionary systems.
\end{abstract}

\maketitle

Error-prone replication, or the copying of genetic sequences with occasional random mutations, is the primary form of genetic variation for asexual, single-celled organisms such as archaea and bacteria~\cite{alberts2022molecular, tippin2004error}, a main thriving mechanism for viruses~\cite{drake1998rates, sniegowski2000evolution, duffy2008rates}, and a source of inspiration for emerging engineering paradigms~\cite{eiben2015introduction, silva2016evolutionary}. It is at the heart of Darwinian evolution, and thought to have played a crucial role before the existence of higher functions such as recombination or horizontal gene transfer~\cite{adamski2020self}.

It was against this backdrop that Manfred Eigen, in the 1970s, introduced his eponymous model~\cite{eigen1971selforganization,eigen1977principle}. The Eigen model (EM) is a prototypical `toy' model of error prone replication, invoking coupled ordinary differential equations (ODEs) to describe how the population fractions of different genetic variants change in time as they replicate, mutate, and deteriorate.
The EM can be solved exactly \cite{saakianExactSolutionEigen2006}, and has proven to be of conceptual importance; notions such as genotype clouds and quasispecies are widely attributed to the EM \cite{eigen1993viral, domingo2012viral, domingo2021historical}. 
It also synonymous with the so-called  `error catastrophe': if mutation rates exceed a
threshold in the EM, the fittest genetic variant--- in the context of the model, the variant with the most rapid replication rate---
does not occupy the largest fraction of the population.

Whilst the original, deterministic formulation of the EM captures the $N\to\infty$ limit, where $N$ is the total population size, several works have since explored the effects of stochasticity due to finite $N$ \cite{alvesErrorThresholdFinite1998, bakhtinEvolutionWeakmutationLimit2021, bonnazStochasticModelEvolving1998, camposFinitesizeScalingError1999, demetriusPolynucleotideEvolutionBranching1985, jonesStochasticAnalysisNonlinear1981, mccaskillStochasticTheoryMacromolecular1984, nowakErrorThresholdsReplication1989, zhangQuasispeciesEvolutionFinite1997,mussoStochasticVersionEigen2011}. These include (but are not limited to) the $N$-dependent scaling of the error threshold \cite{nowakErrorThresholdsReplication1989,alvesErrorThresholdFinite1998,bonnazStochasticModelEvolving1998,camposFinitesizeScalingError1999} and the existence of a dynamics that appears similar to that of `punctuated equilibria' \cite{zhangQuasispeciesEvolutionFinite1997,bakhtinEvolutionWeakmutationLimit2021}.

Notably absent from these works are studies of so-called noise-induced phenomena \cite{NoiseInducedTransitions2006,BiancalaniTKmodel,morrisGrowthinducedBreakingUnbreaking2014,BiancalaniHomoChirality,biancalaniNoiseInducedBistableStates2014,jhawarNoiseinducedSchoolingFish2020}, where finite-$N$
fluctuations are state-dependent and lead to (typically multi-modal)
probability distributions whose extrema do not coincide
with the fixed points of the system’s deterministic, $N\to\infty$ dynamics.
This is surprising not only because such behaviors have been seen in the wider context of evolutionary dynamics--- leading to quasi-cycles \cite{bladon2010evolutionary}, selection `reversal' \cite{constable2016demographic} and non-monotonic rates of demographic switching under population growth \cite{ashcroft2017effects, crosato2023dynamical}--- but also because the stochastic EM and asexual Wright-Fisher (W-F) diffusion have been shown to correspond to different limits of the same upstream stochastic process \cite{musso_relation_2012}. Generically, W-F diffusions are a paradigmatic family of models in population genetics that are characterized by state-dependent fluctuations \cite{james_franklin_crow_introduction_nodate,baake_biological_2000,blythe_stochastic_2007} and therefore noise-induced behavior. The latter has been well-characterized in low dimension, such as when binary mutations are restricted to single locus \cite{james_franklin_crow_introduction_nodate} and corresponds to the so-called weak mutation limit--- where $\mu \ll 1/N$, with $\mu$ the likelihood that a birth gives rise to a mutation.  
The resulting dynamics involves a stochastic switching between two populations that are otherwise effectively clonal--- {\it i.e.}, they behave as if the other variant didn't exist \cite{james_franklin_crow_introduction_nodate}.

Despite such motivation, however, noise-induced behaviour in the (stochastic) EM has not yet been studied. One potential reason for this is its manifestly high-dimensional nature; the number of genetic variants is typically $\sim 4^L$, where $L\gg 1$ is the genome length measured in base pairs. By contrast, and almost without exception, the detailed study of noise-induced effects has seemingly been restricted to cases where the number of variants, or species (dependent on the model) is $\leq 3$. This is because the typical approaches of directly analyzing state-dependent fluctuations--- either visually or algebraically--- are neither illuminating nor straightforwardly tractable in dimension $>3$.

Here, we proceed by calculating the non-dispersive contribution to the probability current of an approximate Fokker-Planck equation (derived via a system-size expansion of the underlying master equation). We then investigate the zeroes of this ``force'' when projected onto a given dimension of phase space. This allows us to identify a critical value for the rate of mutation above which the stationary distribution of the EM is unimodal, and below which it is multimodal. The latter regime, where the fidelity of replication is high, corresponds to a type of stochastic switching between clonallike states. We therefore replace the generic notion of a weak mutation limit with an explicit expression threshold criterion in terms of variant fitness, death rate, system size $N$, and number of variants $m$ ({\it i.e.}, dimension). This enables a surprising observation, in the limit that $m/N\to\infty$, which we might reasonably expect to characterize a large number of real systems, this ``fidelity catastrophe'' threshold coincides with that of the well-known error catastrophe.  That is, the system  either undergoes clonal switching \emph{or} the fittest variant--- the one with the most rapid
replication rate--- acts as a source for all other variants (due to high levels of mutation). In both cases, the fittest variant does not reliably occupy the largest fraction of the population at any given time, subverting standard expectations.

\section*{A Stochastic Eigen Model}

We consider a generalisation of the EM where a population of $N$ genomes is distributed across $m$ variants, with $N_i$ the number of genomes of variant $i\in\{1,\ldots,m\}$.
The genomes replicate and deteriorate at average, variant-specific rates $\alpha_i$ and $\omega_i$, respectively.
During replication, a fraction $\mu_{ij}$ of the new genomes of variant $i$ mutate into variant $j$, on average, with $\hat{\mu}=\mu_{ii}$ the fraction of genomes that do not mutate, which we refer to as the fidelity. 
The following reactions capture the population dynamics
\begin{subequations}%
\begin{align}%
N_i &\xrightarrow{\sum_jN_j\alpha_j\mu_{ji}} N_i + 1,\\
N_i &\xrightarrow{N_i\omega_i} N_i - 1.
\end{align}%
\label{eq:reactions}%
\end{subequations}%
The only restrictions we place on the system, at this stage, are symmetry in the mutation rates, $\mu_{ij}=\mu_{ji}$, and equal fidelity across the variants, $\mu_{ii}=\hat{\mu},\ \forall\,i$.  

In the deterministic, $N\to\infty$, limit Eqs.~(\ref{eq:reactions}) can be approximated by coupled differential equations.
These describe how the population converges, over time, on the demographic mix that corresponds to the largest eigenvalue of the linear algebra problem $W{\bf n}=\lambda{\bf n}$ \cite{saakianExactSolutionEigen2006}, where $W_{ij}\coloneq\alpha_j\mu_{ji}-\omega_i$, and ${\bf n}\coloneq\{n_1,\ldots,n_m\}$ is the demographic mix in concentration space, \emph{i.e.}, $n_i\coloneq N_i/N$. At finite $N$, however, the behaviour of the system significantly deviates from this solution.
%
%
\section*{Noise induced multistability}

Due to the relationship between the stochastic EM and W-F diffusions \cite{musso_relation_2012}, as well as the prevalence of such behavior in evolutionary dynamics \cite{bladon2010evolutionary,constable2016demographic,ashcroft2017effects,crosato2023dynamical}, we are motivated to look for noise-induced bi- or multi-stability.  Noise-induced stability is a concept where state-dependent fluctuations dictate the locations of the extrema ({\it i.e.}, the peaks) of steady-state probability distributions \cite{NoiseInducedTransitions2006}. This typically arises as a consequence of intrinsic fluctuations whose size $\sim 1 / \sqrt{N}$, meaning finite $N$ behavior does not correspond to deterministic, $N\to\infty$ behavior. In systems ranging from idealized spin-systems \cite{morrisGrowthinducedBreakingUnbreaking2014} and molecular autocatalysis \cite{BiancalaniTKmodel,BiancalaniHomoChirality} to collective motion in animals \cite{biancalaniNoiseInducedBistableStates2014,jhawarNoiseinducedSchoolingFish2020}, such behavior is typically marked by a transition from a unimodal distribution, peaked on the deterministic fixed point, to a bimodal distribution.

In the case of a W-F model in one dimension--- {\it i.e.}, binary, single locus--- this switch from uni- to bi-modal is known as the weak mutation limit, which is characterized by $\mu\ll 1/ N$. (Note: despite being broadly adopted elsewhere, the terminology ``noise-induced'' is not used widely in the population genetics literature). More specifically, the peak of the uni-modal case corresponds to a mixed variant fixed-point, whereas the peaks of the bi-modal case correspond to populations where all individuals are from a single variant. As a result, the population switches stochastically between periods of behavior that is otherwise clonal. 

This raises two immediate questions: ({\it i}) does the EM model display noise-induced clonal-switching, and ({\it ii}) can we calculate and/or understand the transition to this regime in a way that is more quantitative that the generic $N\mu\to0$ limit.  

\section*{Effective forces in concentration space}
In order to better understand the role of intrinsic-fluctuations experienced at finite $N$ in the dynamics specified by Eqs.~(\ref{eq:reactions}), we perform a system size expansion~\cite{van1992stochastic, gardiner2009stochastic} (SI~Appendix~I).
This results in a set of It\^{o} stochastic differential equations (SDEs) in $\{\mathbf{n},N\}\coloneq\{n_1,\ldots,n_m,N\}$:
\begin{align}
\label{eq:sde}
\de n_i &= \left(\Phi_i({\bf n}) + \Psi_i({\bf n})\right)\mathop{\de t} + \frac{1}{\sqrt{N}}\sum_{j=1}^m q_{ij}({\bf n})\mathop{\de W_j},\\
\de N&=N(\langle\alpha\rangle_{\mathbf{n}}-\langle\omega\rangle_{\mathbf{n}})\mathop{\de t}+\sqrt{N(\langle\alpha\rangle_{\mathbf{n}}+\langle\omega\rangle_{\mathbf{n}})}\mathop{\de W},
\end{align}%
where
\begin{align}
\Phi_i({\bf n}) &\coloneq A_i({\bf n}) - n_i\sum_{j=1}^m A_j({\bf n}),\\
\Psi_i({\bf n}) &\coloneq -N^{-1}\left(B_i({\bf n}) - n_i\sum_{j=1}^m B_j({\bf n})\right),
\end{align}%
with
\begin{align}
A_i({\bf n})\coloneq\sum_{j=1}^m n_j\alpha_j\mu_{ji} - n_i\omega_i,\\
B_i({\bf n})\coloneq\sum_{j=1}^m n_j\alpha_j\mu_{ji} + n_i\omega_i,
\end{align}%
whilst the multiplicative noise strengths that appear in Eqs.~(\ref{eq:sde}) have the form
\begin{equation}
\label{eq:bdag}
q_{ij}({\bf n}) \coloneq 
\begin{cases}
(1-n_i)\sqrt{B_i(\bf n)} & i=j,\\
- n_i \sqrt{B_j(\bf n)} & i\neq j.
\end{cases}
\end{equation}
In the above, angle brackets $\langle\cdot\rangle_{\mathbf{n}}$ represent the mean with respect to the population fractions, such that $\langle\alpha\rangle_{\mathbf{n}}=\sum_{i=1}^mn_i\alpha_i$ and $\langle\omega\rangle_{\mathbf{n}}=\sum_{i=1}^mn_i\omega_i$. Meanwhile, $\de W_i$ are increments of independent Wiener processes, such that $\mathbb{E}[\de W_i \de W_j]=\delta_{ij}\de t$, where $\delta_{ij}$ is the Kronecker delta, and $\mathbb{E}[\cdot]$ represents an expectation over the noise. We note that the total noise experienced by each $n_i$ is thus highly correlated, originating from the constraint $\sum_{i=1}^m n_i=1$. Finally, the Wiener increment $\de W$ (no subscript) has correlation $\mathbb{E}[\de W \de W_i]=m^{-1/2}\de t$, for all $i$. 
\par
Whilst describing the totality of this system is highly non-trivial, 
we can nevertheless characterise the effect of population size (and hence stochasticity) by studying the instantaneous dynamics for the $n_i$ at arbitrary, fixed $N$. Due the constraint $\sum_{i=1}^m n_i=1$, this restricts us to an $m-1$ dimensional simplex and, as a consequence, the system is over-determined, {\it i.e.}, $p(n_1,\ldots, n_m)=\tilde{p}_{\setminus i}(n_1,\ldots,n_{i-1},n_{i+1},\ldots,n_m)\delta(n_i-(1-\sum_{k\neq i}n_k))$, for all $i$. This is reflected by a non-invertible diffusion matrix  $\mathsf{Q}\coloneq\mathsf{q}\mathsf{q}^T=[Q_{ij}]$, based upon noise strength matrix $\mathsf{q}=[q_{ij}]$.
To circumvent this, we focus on the Fokker-Planck equation (FPE) that corresponds to Eqs.~(\ref{eq:sde}), but, without loss of generality, with the $m$-th dimension removed (SI~Appendix~II).
This leads to a reduced FPE which can be written
\begin{align}
\label{eq:fpe-minus}
\dot{\tilde{p}}(\tilde{{\bf n}}, t)= &-\sum_{i=1}^{m-1}\partial_{n_i}\Bigg[\Big(\Phi_i(\tilde{\mathbf{n}})+\Psi_i(\tilde{\mathbf{n}})+\Theta_i(\tilde{\mathbf{n}})\Big)\tilde{p}(\tilde{\mathbf{n}}, t)\nonumber\\
&\qquad\qquad\quad-\frac{1}{2N}\sum_{j=1}^{m-1}\mathcal{Q}_{ij}(\tilde{\mathbf{n}})\partial_{n_j}\tilde{p}(\tilde{\mathbf{n}}, t)\Bigg].
\end{align}
Here $\tilde{p}(\tilde{{\bf n}}, t)\equiv \tilde{p}_{\setminus i}$ is the probability density over population densities $\tilde{\mathbf{n}}=\{n_1,\ldots,n_{m-1}\}$, whilst $\mathcal{Q}_{ij}$ are elements of the invertible sub-matrix $\mathcal{Q}\coloneq [\mathsf{Q}]_{1:m-1;1:m-1}=[\mathcal{Q}_{ij}]$.  
Meanwhile, the $\Theta_i$ term is given by
\begin{equation}
\Theta_i(\tilde{\mathbf{n}}) \coloneq -\frac{1}{2N}\sum_{j=1}^{m-1}\partial_{n_j} \mathcal{Q}_{ij}(\tilde{\mathbf{n}}),
\label{eq:noise-ind-term}
\end{equation}
 which we refer to as the \emph{noise-induced effective force}. 
Importantly, for Eq.~(\ref{eq:fpe-minus}) to be a closed equation in terms of $\tilde{\mathbf{n}}$ we must have every appearance of $n_m$ replaced by $1-\sum_{j=1}^{m-1}n_j$, for example: $\Phi_i(\tilde{\mathbf{n}})=\Phi_i(\mathbf{n})|_{n_m=1-\sum_{j=1}^{m-1} n_j}$. 
\par
The presentation of Eq.~(\ref{eq:fpe-minus}) deliberately casts the total flux--- {\it i.e.} the terms inside the square brackets--- as being formed from both an advective term, or \emph{effective force}, which we write  $F_i(\tilde{\mathbf{n}}) = \Phi_i(\tilde{\mathbf{n}})+\Psi_i(\tilde{\mathbf{n}})+\Theta_i(\tilde{\mathbf{n}})$, and a remaining dispersive component. 
The meaning of such contributions become clear if we consider the limiting distribution ($\dot{\tilde{p}}=0$), under the assumption of zero stationary flux, such that we can write, in vector form,
\begin{equation}
\label{eq:st-condition}
\frac{\nabla \tilde{p}(\tilde{\mathbf{n}})}{\tilde{p}(\tilde{\mathbf{n}})} = 2N \mathcal{Q}^{-1}(\tilde{\mathbf{n}})F(\tilde{\mathbf{n}}).
\end{equation}
Here $\nabla=(\partial_{n_1},\ldots,\partial_{m-1})$, and $\Phi(\tilde{\mathbf{n}})=(\Phi_1(\tilde{\mathbf{n}}),\ldots,\Phi_{m-1}(\tilde{\mathbf{n}}))$ \emph{etc.}, such that $F(\tilde{\mathbf{n}}) = \Phi(\tilde{\mathbf{n}})+\Psi(\tilde{\mathbf{n}})+\Theta(\tilde{\mathbf{n}})$ represents the vector of the \emph{total effective force}, which is proportional to the advective part of the total flux. Such a quantity can be used as a proxy for the expected behaviour, or drift, of the system in any given population state, $\tilde{\mathbf{n}}$. For example, in the above case, where the total flux vanishes in the stationary state, extrema in $\tilde{p}(\tilde{\mathbf{n}})$ coincide with the zeros of $F(\tilde{\mathbf{n}})$. These therefore act as effective dynamical fixed points, which may be stable or unstable if they correspond to maxima and minima in the distribution, respectively.

Generically, however, we cannot guarantee that the stationary flux vanishes in the (stochastic) EM. We nevertheless conjecture that transition to switching behaviour--- {\it i.e.} from unimodality to multimodality--- will be captured by the appearance (or disappearance) of zeros in $F(\tilde{\mathbf{n}})$. If true, this weaker property is still helpful because, whilst solution of the FPE is, in general, not possible, we can readily derive $F(\tilde{\mathbf{n}})$ (given in full in  SI Appendix~III). 
\par
Finally, we note that in the case of a single dimension, the division of the current in Eq.~(\ref{eq:fpe-minus}) and the appearance of the noise induced drift can be associated with conversion of the underlying SDEs between an It\^{o} and a H{\"a}angi-Klimontovic interpretation \cite{pacheco-pozoLangevinEquationHeterogeneous2024,sokolovItoStratonovichHanggi2010}. However, in higher dimension with correlated noise, as considered here, this property no longer holds.

\section*{Projection onto simplex bisection lines}

\begin{figure}[t]
    \centering
    \includegraphics[width=\columnwidth]{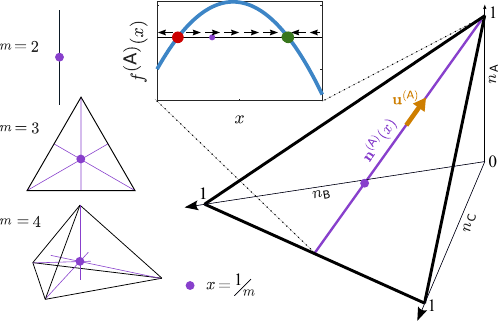}
    \caption{Cartoon of force projection and inspection. On the left, three simplexes are shown for the cases of $m=\{2,3,4\}$. All bisecting lines $\mathbf{n}^{(c)}(x)$ are shown in purple, highlighting the center of the simplex, where $n_i=n_j \forall i,j$ and corresponding to $x=1/m$, where all bisecting lines meet. On the right, the case of $m=3$ is considered more in detail. The three variants are labeled $\textsf{A}$, $\textsf{B}$ and $\textsf{C}$, with the purple line representing $\mathbf{n}^{(\textsf{A})}(x)$, bisecting the \textsf{A}-corner and the center of the simplex. Note that $\mathbf{n}^{(\textsf{A})}(0)$ corresponds to a population with no \textsf{A}-genomes, while $\mathbf{n}^{(\textsf{A})}(1)$, \emph{i.e.}, the \textsf{A}-corner itself, to a population of only \textsf{A}-genomes. The orange arrow, arbitrarily positioned along the bisecting line, represents $\mathbf{u}^{(\textsf{A})}$ (not to scale). Finally, the inset illustrates how projected forces over $x$ are visualized.}
    \label{fig:projection}
\end{figure}

Since $F(\tilde{\mathbf{n}})$ is a vector field in $m-1$ dimensions, with $m$ potentially very large, its investigation is both visually and analytically complicated (SI~Appendix~III).
To simplify the analysis, we choose to inspect $F(\tilde{\mathbf{n}})$ along lines that bisect the centre of the simplex and its vertices, characterised by co-ordinates $\mathbf{n}^{(c)}(x)=(n_1^{(c)}(x),\ldots, n_m^{(c)}(x))$, $x\in[0,1]$, with
\begin{equation}
n_i^{(c)}(x)=
\begin{cases}
x & i=c,\\
\frac{1-x}{m-1} & i\neq c,
\end{cases}
\label{eq:nc}
\end{equation}
where $c$ is a specifically chosen co-ordinate determining the bisection line (example in Fig.~\ref{fig:projection}). 
Explicitly, $\mathbf{n}^{(c)}(1)$ corresponds to a system comprised only of variant $c$, whilst $\mathbf{n}^{(c)}(0)$ corresponds to a system completely absent variant $c$, and with all other variants present in equal proportion. 
\par
Moreover, we not only inspect $F(\tilde{\mathbf{n}})$ on the line $\mathbf{n}^{(c)}(x)$, but also \emph{in the direction} of this line, which is given by the unit vector ${\bf u}^{(c)}$ pointing towards the corner $\mathbf{n}^{(c)}(1)$ of the simplex:
\begin{equation}
u_i^{(c)}=
\begin{cases}
\sqrt{\frac{m-1}{m}} & i=c,\\
-\sqrt{\frac{m-1}{m}}\frac{1}{m-1} & i\neq c.
\end{cases}
\label{eq:u}
\end{equation}
To characterise the effective force along the simplex bisection line, we are ultimately interested in the scalar projection
\begin{equation}
f^{(c)}(x) = F(\mathbf{n}^{(c)}(x))\cdot {\bf u}^{(c)} = \sum_i F_i(\mathbf{n}^{(c)}(x))u^{(c)}_i,
\label{eq:scalar-proj-tot}
\end{equation}
and its partition into $f^{(c)}(x) = f^{(c)}_\Phi(x) + f^{(c)}_\Psi(x) + f^{(c)}_\Theta(x)$.

It is then instructive to consider the effective 1D dynamics due to $f^{(c)}(x)$. To do so, we calculate exact expressions for each of the three contributions--- $f^{(c)}_\Phi(x)$, $f^{(c)}_\Psi(x)$, and $f^{(c)}_\Theta(x)$--- in the general case, in SI~Appendix~IV. Crucially, all such components are quadratic functions of the parameterization variable, $x$. 
This means that there can be up to two zeroes in the projected effective force within the relevant interval, $[0,1]$. That is, up to two fixed points in the effective dynamics. In the case of stationary flux, such fixed points can be identified with the extrema of the stationary distribution: a stable fixed point corresponds to a maximum and an unstable fixed point a minimum. Whilst we cannot formally make this identification here, due to the absence of stationary flux, we assert (and later verify) that a \emph{change} in the number of zeroes within $[0,1]$ will still be indicative of a change in the number of peaks in the stationary distribution.  

\begin{figure*}[t]
    \centering
    \includegraphics[width=\textwidth]{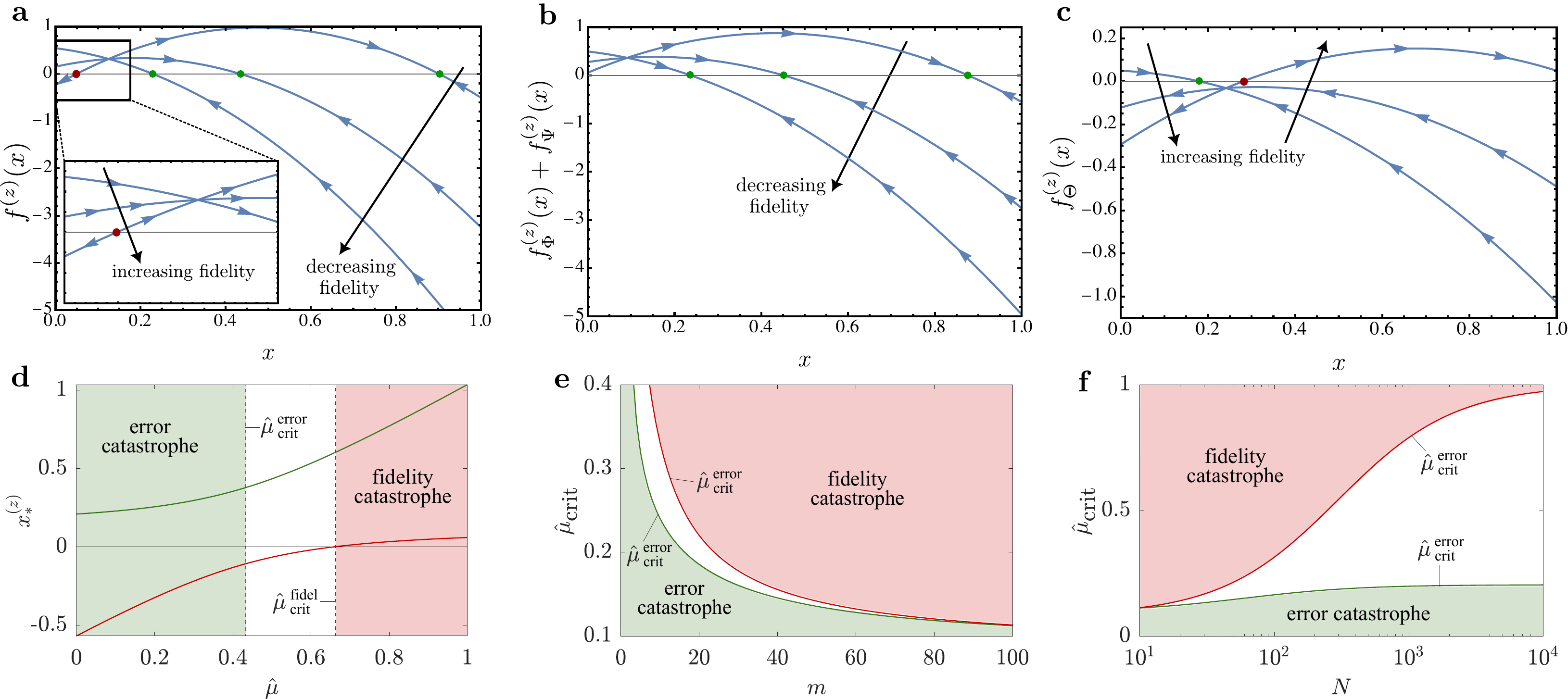}
    \caption{Effective force along the simplex bisection line of the WT variant.
    Across all panels $\alpha=1$, $\Delta\alpha=4$, and $\omega=0.5$.
    \textbf{a}) Setting $m=3$ and $N=10$, the blue lines show the total effective force, $f^\textrm{(z)}(x)$, at different fidelities, $\hat{\mu}$, taking values  $0.1$, $0.5$, and $0.9$.
    As the fidelity is increased the lower fixed point (red dot) enters the simplex such that the force at $x=0$ becomes negative.
    On the other hand, as the fidelity is reduced (error is increased), the upper, stable, fixed point (green dots) moves to lower values of $x$ in the simplex.
    \textbf{b}) As panel \textbf{a}), but showing only the deterministic component of the effective force --- there is no unstable fixed point within the simplex for any $\hat{\mu}$. 
    \textbf{c}) As panel \textbf{a}), but showing only the noise-induced component of the effective force --- for high $\hat{\mu}$ an unstable fixed point arises. For suitably low $N$ this can contribute sufficiently to the total force that such a fixed point appears in the total force also (\emph{cf.} panel \textbf{a}))
    \textbf{d}) With again $m=3$ and $N=10$, the green and red lines show the position of the stable and unstable fixed point, respectively, as a function of $\hat{\mu}$.
    The critical values of $\hat{\mu}$ marking the onset of the error catastrophe and weak mutation regime are highlighted by dashed lines.
    The onset of the weak mutation regime is defined as the point where the unstable fixed point enters the simplex at $x=0$, while the onset of the error catastrophe is defined by the point of maximum curvature in the location of the stable fixed point with $\hat{\mu}$.
    \textbf{e}) Setting $N=10$, the positions of the thresholds are shown as a function of $m$.
    \textbf{f}) Setting $m=100$, the positions of the thresholds are shown as a function of $N$.}
    \label{fig:fixed-points}
\end{figure*}

\section*{Application to a model of Quasispecies}

To better illustrate our arguments, consider a common scenario in the study of quasispecies \cite{camposFinitesizeScalingError1999,alvesErrorThresholdFinite1998}: a single dominant `wild-type' (WT) variant, $z$, that replicates faster than all other variants, such that $\omega_i=\omega$ $\forall i$, $\alpha_i=\alpha$ $\forall i\neq z$, $\alpha_z=\alpha + \Delta\alpha$, with $\Delta\alpha>0$. 
Whilst an analysis of non-WT variants is provided in SI Appendix~VI, the salient properties of the system are captured by considering just the force projection onto the WT's bisection line  (\emph{i.e.} $c=z$).
In this case, the effective force is a negative quadratic in $x$, illustrated qualitatively in Fig.~\ref{fig:projection}. Generically, the upper fixed point is stable, whilst the lower fixed point is unstable.  However, crucially, the position of these fixed points--- including whether they lie inside or outside of the simplex--- varies with system parameters. In particular, the variation with the fidelity ($\hat{\mu}=\mu_{ii}$) is illustrated in Fig.~\ref{fig:fixed-points}\textbf{a}. Here, as the fidelity of the system is increased, the stable fixed point moves towards $x=1$ (purely WT population) as might be expected. However, this also moves the unstable fixed point in the same direction. At high enough values of $\hat{\mu}$ (or conversely at low enough values of $N$) the unstable fixed point enters the interval $[0,1]$. This can be attributed to the action of the noise-induced component of the effective projected force, $f^{(z)}_\Theta(x)$. When this component is removed, such as in the deterministic $N\to\infty$ limit, the unstable fixed point does not enter the simplex, irrespective of $\hat{\mu}$ (Figs.~\ref{fig:fixed-points}\textbf{b} \& \textbf{c}).
\par
Notably, when the unstable fixed point lies within the simplex, a qualitatively new dynamics can take place: any state whose projection lies between that of the unstable fixed point and $x=0$ will be driven towards $x=0$ (and not $x=1$). That is, if the system comprises only a small fraction of WTs, it will be driven towards states where there are none, despite the WT being the fittest variant.
In terms of the the probability distribution, this corresponds to the appearance of local minima, causing an otherwise unimodal distribution to become multimodal. 

Such behavior is ultimately the result of a simple positive feedback process. To see this, consider a random fluctuation that leads to an arbitrary non-WT variant possessing a larger population fraction than all the others. When $N$ is large, there is an overwhelming likelihood that this difference is small, and the system will respond by reverting to the mean. However, when $N$ is small, such a fluctuation can lead to substantive differences in population fractions. Here, since there is a fixed birth rate {\it per individual}, more of that variant are likely to be produced, relative to the others, causing it to possess a yet larger fraction of the population. This feedback leads to the temporary fixation of the random variant--- or so-called clonal behavior--- and therefore depletion of the WT. Subsequent fluctuations, however, can then start the process all over again, leading to intermittent stochastic switching between periods of clonallike behavior of random single variants.

Of note, the way this mechanism manifests mathematically, in a small $N$ stochastic description in terms of population fractions, is via fluctuations, or noise, that is state-dependent. To see this, consider a state where a single variant is dominant--- {\it e.g.,} ${\bf n} = \{1,0,\ldots,0\}$. Here, the stochastic effects of birth and death do not lead to fluctuations. By contrast, in a fully mixed state--- {\it e.g.,} ${\bf n} = \{N/m,N/m,\ldots,N/m\}$--- each birth and death event changes the relative composition of the system and hence generates fluctuations. This gradient in fluctuations, from mixed states ({\it i.e.}, the center of the simplex) to single variant states ({\it i.e.}, the corners of the simplex), results in noise-induced forces that codify the above feedback mechanism in terms of a multimodal distribution peaked on single-variant states.

\section*{A threshold for the fidelity catastrophe}

Clonal switching will only take place under certain conditions. If the fidelity is too low, replication can be insufficiently clonal, being more likely to produce distinct variants, interrupting the positive feedback, and instead causing the population distribution to be more mixed. On the other hand, if $N$ is too large fluctuations become too small and the stochastic effect is overwhelmed by the deterministic ($N\to \infty$) drift of the system.
\par
In particular, because we can identify the onset of this behavior in terms of the entry of the unstable fixed point into the simplex, we can characterize the critical system parameters that act as threshold values to the fidelity catastrophe.
Specifically, the fixed point enters the simplex when $f^{(c)}(x=0)=0$ such that it lies exactly on the edge of the simplex. By solving for this condition in terms of $\hat{\mu}$, we thus determine the critical fidelity 
above which the distribution becomes multimodal, and below which the distribution remains unimodal, which we calculate to be (see SI Appendix.~VI)
\begin{align}
\label{eq:crit-mu}
\hat{\mu}^{\rm fidel}_{\rm crit}&=\frac{\alpha  (m+2 N-1)-m \omega +\omega }{\Delta\alpha  (m-1)+2 \alpha  (m+N-1)}.
\end{align}
Equivalently, we may consider this behavior to occur at a critical $N$, at fixed $\hat{\mu}$, in the same manner, yielding
\begin{align}
\label{eq:crit-N}
N^{\rm fidel}_{\rm crit}&=\frac{(m-1) (\alpha  (2 \hat{\mu} -1)+\Delta\alpha \hat{\mu} +\omega )}{2 \alpha  (1-\hat{\mu})},
\end{align}
above which the distribution is unimodal, and below which it is multimodal. 
In particular, we note that $N^{\rm fidel}_{\rm crit} \sim m$. In other words, the sense in which $N$ can be considered small, such that there are noise-induced effects, grows with the dimensionality of the system.

\section*{Revisiting the error catastrophe}

Given that the Eigen model is near synonymous with the so-called error catastrophe, or error threshold, wherein the most fit variant is no longer the most numerous variant above a threshold error (or below a threshold fidelity), it is instructive to revisit such a phenomenon in the context of our stochastic model. 
Since our model is stochastic, identifying such a threshold necessarily involves a generalization of such a concept.  However, by studying the properties of the fixed points, we find that we can offer a meaningful definition of such a threshold that recovers the classical result in the appropriate limit.
\par
In particular, we expand on the observation that the location of the stable fixed point moves through the simplex as the error/fidelity is varied (\emph{cf.} Fig.~\ref{fig:fixed-points}\textbf{a}). More specifically, we observe that as $\hat{\mu}$ is reduced, the system can experience a rapid change between a regime of approximately linear response in the position of the stable fixed point and a regime where it is relatively independent of $\hat{\mu}$ and lies near $x=1/m$ (see SI Appendix~VI). Moreover, as $m$ is taken larger this transition becomes sharp. Consequently, we choose to characterize a threshold between these behaviors through the point of maximal curvature in the position of the fixed point $x^\textrm{(z)}_*$ with respect to $\hat{\mu}$, \emph{i.e.}, $\de^3 x^\textrm{(z)}_*(\hat{\mu})/\de\hat{\mu}^3=0$. Since the location of the fixed point is the solution of a quadratic this has an analytical solution, occurring at a critical fidelity, $\hat{\mu}^{\rm error}_{\rm crit}$, reported in SI Appendix~VI. Such an expression amounts to a powerful generalization of the original heuristic formula for the error threshold, containing both population and phase space/genome size dependence through $N$ and $m$, respectively. 
\par
Notably, the error threshold places the opposite bound on the system in terms of the fidelity, $\hat{\mu}$, as that of the fidelity catastrophe, in that for the error catastrophe to be avoided we must have $\hat{\mu}\geq\hat{\mu}^{\rm error}_{\rm crit}$, whilst for the fidelity catastrophe, mutimodal, regime to be avoided we must obey $\hat{\mu}\leq\hat{\mu}^{\rm fidel}_{\rm crit}$. The result of this is that there is a characteristic `window', where $\hat{\mu}^{\rm error}_{\rm crit}\leq \hat{\mu}\leq\hat{\mu}^{\rm fidel}_{\rm crit}$, in which the system avoids both scenarios. This is illustrated in Fig.~\ref{fig:fixed-points}\textbf{d}-\textbf{f} where the behavior of the two fixed points, and their relationship to the two thresholds, is clearly demonstrated.
\par
We note that, despite the dependence of our $\hat{\mu}^{\rm error}_{\rm crit}$ on $m$ and $N$, the error threshold originally postulated by Eigen is easily recovered by progressively considering the deterministic ($N\to \infty)$, and then large state space ($m\to\infty$), limits, obtaining $\alpha/(\alpha+\Delta \alpha)$ for our quasispecies example \cite{eigen1971selforganization,eigen1977principle}.
\par
Interestingly, however, the location of this threshold changes if we consider the limits to be taken in the opposite order, such that we consider $m\gg N$. In such a regime the generalized error threshold converges on $(\alpha-\omega)/(2\alpha+\Delta\alpha)$. Moreover, this is precisely the value that Eq.~(\ref{eq:crit-mu}) takes in the same limit. Explicitly, in the limit $m\gg N$ the thresholds to the error and fidelity catastrophes converge to the same value, with their ratio obeying 
\begin{align}
\frac{\hat{\mu}_{\rm crit}^{\rm fidel}}{\hat{\mu}_{\rm crit}^{\rm error}}&=1+\frac{2\Delta\alpha}{2\alpha+\Delta\alpha}\frac{N}{m}+\mathcal{O}(m^{-1}).
\label{eq:error_ratio}
\end{align}
This suggests that the `window' of possible $\hat{\mu}$ for which both the error catastrophe and fidelity catastrophe are avoided vanishes in this limit. In other words, in our quasispecies example, should the number of variants far exceed the number of replicators, in order to avoid the error catastrophe, one must be in the fidelity catastrophe regime such that the distribution is multimodal.

\begin{figure*}[!]
\centering
\includegraphics[width=\textwidth]{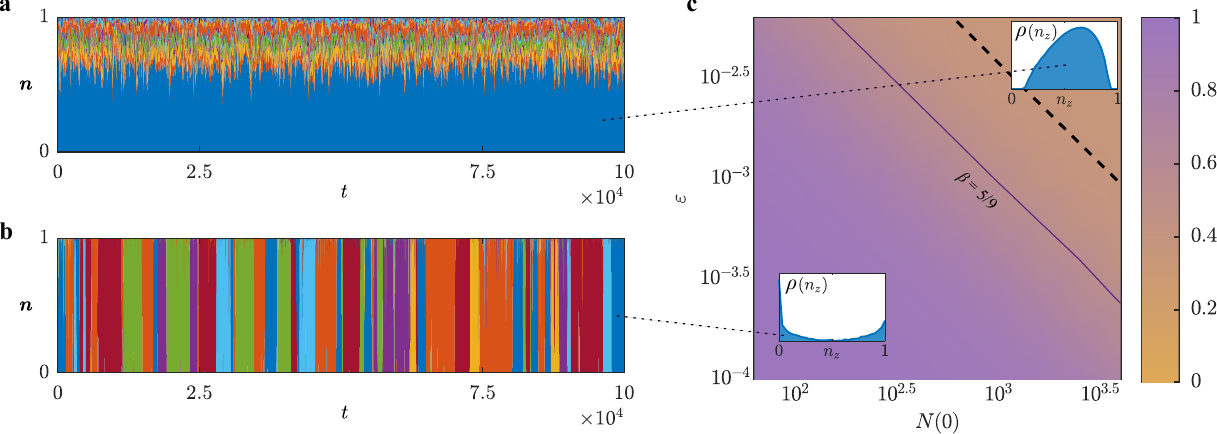}
\caption{Numerical evidence for the transition to the weak mutation regime.
The following setup is considered: $|{\cal A}|=2$, $L=4$, $\omega_i=1$ $\forall i$, $\alpha_i=1$ $\forall i\neq z$, $\alpha_z=1+10\varepsilon$ and $\mu_{ij}$ as per Eq.~(\ref{eq:mut-ham}).
The dynamics in Eqs.~(\ref{eq:reactions}), coupled with a saturation constraint, are simulated for different combinations of $\varepsilon$ and $N$.
\textbf{a}) Time evolution of the concentrations ${\bf n}$ with $N(0)=10^{3.2}$ and $\varepsilon=10^{-2.5}$, each color representing one of the 16 variants with the WT dark blue.
\textbf{b}) Time evolution of the concentrations ${\bf n}$ with $N=10^{2}$ and $\varepsilon=10^{-3.8}$ (colors as in panel \textbf{a}).
\textbf{c}) Bimodality coefficient of $n_z$ over time, $\beta = (\textrm{skew}[\rho(n_z,t)]^2+1)/\textrm{kurt}[\rho(n_z,t)]$.
The two insets are examples of distributions of $n_z$ (note the y-axis is logarithmic), corresponding to the two set of parameters of panels \textbf{a} and \textbf{b}.
The purple line marks the contour where $\beta=5/9$, the threshold above which the distribution is to be considered bimodal.
The black dashed line indicates the onset of the weak mutation regime, as per Eqs.~(\ref{eq:crit-mu}) and~(\ref{eq:crit-N}).}
\label{fig:quasispecies}
\end{figure*}

\section*{Population size and dimensionality}

In light of Eq.~(\ref{eq:error_ratio}), we consider the importance of the ratio $N/m$, and the extent to which these quantities are naturally antagonistic when discussing noise induced effects.
\par
We start by considering the effect of varying $m$ at fixed $N$ on the position of the fidelity thresholds, with the dependence illustrated in Fig.~\ref{fig:fixed-points}\textbf{e}.
As $m$ increases, both thresholds move to lower values of $\hat{\mu}$, and we can clearly see the prediction of Eq.~(\ref{eq:error_ratio}): the two thresholds tend to each other, reducing the fidelity interval where neither the error catastrophe or fidelity catastrophe behavior is present.
By contrast, the effect of varying $N$ at fixed $m$ is shown in Fig.~\ref{fig:fixed-points}\textbf{f}.
Here, as $N$ increases, the position of the two thresholds moves to higher values of $\hat{\mu}$, as they also tend away from each other, thus increasing the fidelity interval where neither behavior is present.
Here we can also appreciate that the fidelity catastrophe is clearly a stochastic effect, with larger and larger fidelities (tending to $1$) required to observe the effect as $N$ is taken larger.
\par
In totality, the central observation that underlies the behavior is the importance of the ratio $N/m$. In particular, we argue that it refines the idea of what constitutes a `low $N$' system, wherein stochastic effects, such as noise induced multi-stability, become non-negligible.
Crucially, when this ratio is large (low-$m$ side of Fig.~\ref{fig:fixed-points}\textbf{e} and high-$N$ side of Fig.~\ref{fig:fixed-points}\textbf{f}) there are a large number of replicators \emph{per variant} and thus stochastic effects are small.
In this limit the fidelity catastrophe behavior only occurs for extremely large values of $\hat{\mu}$ ($\sim 1$) and generally does not occur in the vicinity of the error catastrophe.
By contrast, when $N/m$ is small (high-$m$ side of Fig.~\ref{fig:fixed-points}\textbf{e} and low-$N$ side of Fig.~\ref{fig:fixed-points}\textbf{f}) there are, on average, very few replicators \emph{per variant} (\emph{i.e.}, significantly less than one) and thus stochastic effects are highly significant. Indeed, we can directly quantify the critical ratio, $\xi\coloneq N/m$, below which we can expect noise induced multi-stability by inspection of Eq.~(\ref{eq:crit-N}), giving
\begin{align}
    \xi^{\rm fidel}_{\rm crit}&=\frac{m-1}{m}\frac{(\alpha  (2 \hat{\mu} -1)+\Delta \alpha  \hat{\mu} +\omega )}{2 \alpha  (1-\hat{\mu})}
\end{align}
which becomes independent of $N$ and $m$, as $m\to \infty$, such that it is solely the ratio, $\xi$, which determines whether noise induced effects arise. That is we have
\begin{align}
    \lim_{m\to \infty}\hat{\mu}^{\rm fidel}_{\rm crit} = \frac{2 \alpha  \xi +\alpha -\omega }{2 \alpha  (\xi +1)+\Delta \alpha }.
\end{align}
\par
It is then not irrelevant that for realistic models of genomes and populations, the number of variants $m$ will be extremely large (much larger than $N$) such that $\xi\ll 1$ and fluctuations become important. When interpreted at face value this implies that realistic systems cannot avoid both the error catastrophe and the fidelity catastrophe simultaneously. 

\section*{Numerical evidence}

In order to test our predictions and illustrate the fluctuating nature of the fidelity catastrophe we simulate the system described by Eqs.~(\ref{eq:reactions}) using the Gillespie algorithm \cite{gillespieGeneralMethodNumerically1976}. As with the set-up described previously, which underpins Eqs.~(\ref{eq:crit-mu}) and (\ref{eq:crit-N}), we consider the scenario of equally fit variants except the WT which replicates faster.

A realistic genetic model is assumed where genomes consist of $L$ bases in the set ${\cal A}$ ({\it e.g.}, ${\cal A}=\{\textrm{A,T,G,C}\}$), leading to a total of $m=|{\cal A}|^L$ variants connected by a mutation matrix, $\mu=[\mu_{ij}]$, with elements
\begin{equation}
\label{eq:mut-ham}
\mu_{ij}=\left(\frac{\varepsilon}{|{\cal A}|-1}\right)^{H_{ij}}(1 - \varepsilon)^{L - H_{ij}},
\end{equation}
where $\varepsilon$ is the average error rate per base during replication and $H_{ij}$ is the Hamming distance, {\it i.e.}, the amount of loci where the genomes of $i$ and $j$ have different bases. 
Note, the per base error and fidelity are thus related by $\hat{\mu}=(1-\varepsilon)^L$. 
In order to characterise the system at a given $N$, the population size is kept stable around its initial value $N(0)$ by adding a a saturation limit (see SI Appendix~VIII). Further, we make the selective advantage, $\Delta \alpha$, a function of $\varepsilon$ (SI Appendix~VIII and Fig.~\ref{fig:quasispecies}), so that the deterministic solution (\emph{i.e.}, the composition of dominant quasispecies), which depends on both $\varepsilon$ and $\Delta \alpha$, is comparable across all simulations.

As numerical simulation is only possible for relatively small numbers of variants, we start by considering a population of just $m=16$ variants.
We explore the behavior of this population at various combinations of population size and error rate (Fig.~\ref{fig:quasispecies}), in a regime where the error catastrophe is avoided.
Two drastically different behaviors are observed: either the concentrations fluctuate over time around the deterministic solution (Fig.~\ref{fig:quasispecies}\textbf{a}), or individual variants randomly take turns at dominating the entire population for long stretches of time, regardless of their Hamming distance to the WT (Fig.~\ref{fig:quasispecies}\textbf{b}).
The distribution of the concentration of the WT over time, $\rho(n_z, t)$, identifies the two regimes corresponding to observed behaviors (Fig.~\ref{fig:quasispecies}\textbf{a}): at higher error rate (low fidelity) and larger population size, $\rho(n_z,t)$ is unimodal and peaked on the value predicted by the deterministic model, while at lower error rate (high fidelity) and smaller population size, $\rho(n_z,t)$ is bimodal with peaks at $n_z=0$ and $n_z=1$. Such behavior well characterizes a system within, and outside, the fidelity catastrophe regime, respectively.
\par
To test the predictions of our theory, we substitute $\hat{\mu}$ with $\varepsilon$ in either Eq.~(\ref{eq:crit-mu}) or Eq.~(\ref{eq:crit-N}) to obtain a line which bounds the weak mutation regime. The result yields the dashed black line in Fig.~\ref{fig:quasispecies}\textbf{c}, whilst the solid purple line marks a nominal demarcation of unimodal and multimodal empirical probability distributions. We observe that that our prediction well captures the trend of the transition between unimodality and bimodality, with weak mutation switching dynamics occurring at large $\hat{\mu}$ (low $\varepsilon$) where the unstable fixed points resides in the simplex. However, quantitatively, its position is overestimated compared to the nominal unimodal/multimodal line. Such a discrepancy could be attributed to several factors including limitations of the system size expansion that underlies Eq.~(\ref{eq:crit-mu}), or the necessity for the unstable fixed point to be farther into the simplex for its effects to be noticeable.

\section*{Discussion and conclusions}

We have studied the stochastic EM in high dimension with a focus on the noise-induced behavior that manifests when replication occurs with high fidelity ({\it i.e.}, mutation is weak). In particular, by using an effective force projection method, we have been able to characterize the qualitative behavior of the system through the description of two emergent fixed points. One of these fixed points is stable and associated with the deterministic behavior of the system. It allows us to characterize an error threshold, by analogy with the error catastrophe that is synonymous with the deterministic version of the EM. The other fixed point is unstable and associated with noise induced effects. It influences behavior at finite $N$ only, with its entry into the simplex defining another threshold, this time at high fidelity, marking the entry to a regime where solutions become multimodal.
\par
By calculating these thresholds precisely, we are able to determine how they vary with system parameters. The dependence on the dimension of the system, $m$, is particularly striking. Specifically, we observe that $m$ plays an antagonistic role with system size, whereby high dimensional systems need a commensurately large number of replicators, $N$, in order to avoid noise induced behaviors from manifesting. Consequently, it is the ratio $N/m$, \emph{i.e.} the number of replicators \emph{per variant}, which controls the onset of noise induced multistability in systems with arbitrary dimension.
\par
In light of this, it is highly relevant that many realistic scenarios will have very large dimension, such that $m\gg N$ (typically $m\sim4^L$, with $L\gg 1$ the length of the genome). In this regime we have shown that the threshold to the fidelity catastrophe and a generalized error threshold coincide. The implication is that, assuming that the error threshold is to be avoided, these systems will generically be multistable and therefore exhibit noise induced behaviors.
\par
This result is surprising and, at first take, contrary to all standard expectations of evolutionary systems. However, whilst we have demonstrated a threshold for the existence of mutlistablity, we have stopped short of quantifying its dynamical aspects; a detailed analysis of the effective transition rates between clonallike states is required to properly capture the relative likelihood of visiting, say, the master variant, as well as how long the system remains, clonally, in that state. For example, in cases such as those with large selection advantages, it may be that whilst a distribution is technically multimodal, one must wait for an extremely long time to observe a stochastic fluctuation away from the WT. Precise expressions for the transition rates are likely to rely on the magnitude by which the replication rate of the fittest variant exceeds that of the other variants ({\it i.e.}, $\Delta\alpha$ in our quasispecies example), the number of variants ({\it i.e.}, the dimension, $m$) and the underlying structure of the mutation rates ({\it i.e.}, a chosen Hamming or other model). Recently, Piecewise Deterministic Markov Processes have been used to approximate  transitions in stochastically multi-stable systems and this technique may also prove useful here \cite{hufton_intrinsic_2016,hufton_model_2019,worsfold_stochastically_2025}.
\par
There are also other limitations of our analysis that further work may help to resolve. Whilst we have presented effective force calculations for arbitrary fitness landscapes, we have restricted ourselves to the specific case of a single peaked landscape for the purpose of calculating explicit thresholds. In practice, fitness landscapes can be expected to be far more complex. Moreover, we remark that the Eigen model is concerned with pointwise, or single locus mutations.  That is, every distinct genome is a viable variant. In reality, however, there is significant redundancy in the mapping between mutants and fitness: many single locus mutations do not confer changes in fitness, which typically require cumulative mutation. As a result, the effective dimension, in terms of distinct phenotypes, is significantly smaller than $m$, challenging the notion that $m/N$ is a crucial parameter. An interesting question therefore appears to be: how much of the Eigen model's phenomenology survives under a phenotypic coarse-graining? We therefore welcome all further work in the area.

\section*{Declarations}
The authors have no competing interests to declare.

\begin{acknowledgments}
EC, RS and RGM acknowledge support from the EMBL Australia program. RGM acknowledges funding from the Australian Research Council Centre of Excellence for Mathematical Analysis of Cellular Systems (CE230100001). The authors thank K.~Husain (UCL, London, U.~K.) for critical feedback on initial versions of the manuscript.
\end{acknowledgments}

\clearpage
\onecolumngrid
\ 
\vspace{2.2cm}
\begin{center}
\textbf{\large{SUPPLEMENTARY INFORMATION}}
\end{center}
\vspace{2.2cm}
\renewcommand{\thesection}{\Roman{section}}

\section{Formulation of the stochastic dynamics}
\label{sec:sse}

Let us denote by $p({\bf N},t)$ the probability of a specific demographic mix ${\bf N}\coloneq\{N_1,\ldots,N_m\}$ at time $t$.
The stochastic dynamics of the system are captured by the master equation
\begin{equation}
\dot{p}({\bf N},t) = \Bigg[
\sum_{i,j=1}^m\left(E_{N_i}^{-1} - 1\right)N_j\alpha_j\mu_{ji}
+\sum_{i=1}^m \left(E_{N_i}^{+1}-1\right)N_i\omega_i\Bigg] p({\bf N},t),
\label{eq:master}
\end{equation}
where the step operator $E_x^r$ has action $E_x^r f(x)=f(x+r)$. We recall that $\alpha_i$ and $\omega_i$ are the replication and deterioration rates of variant $i$, respectively, whilst $\mu_{ji}$ is the probability that variant $i$ is produced upon the replication of variant $j$, such that $\sum_{i=1}^m\mu_{ji}=1$. As such, we consider $1-\mu_{ii}$ to be the total replication error of variant $i$, whilst its complement, $\mu_{ii}\equiv \hat{\mu}$, is its fidelity.
\par
A set of Stochastic Differential Equations (SDEs) that approximate the dynamics  can be obtained from the master equation through a system size expansion~\cite{van1992stochastic}.
When $r\ll x$, the step operator $E$ can be approximated using a Taylor series up to the second order, \emph{i.e.},
\begin{equation}
E^r_x -1 \approx r\frac{\partial}{\partial x} +\frac{r^2}{2}\frac{\partial^2}{\partial x^2}.
\label{eq:op-apprx}
\end{equation}
Hence, for large values of $N\coloneq\sum_{i=1}^m N_i$, we can substitute Eqs.~\eqref{eq:op-apprx} into~\eqref{eq:master} to obtain the Fokker-Plank (FP) equation
\begin{equation}
\begin{aligned}
\frac{\partial p({\bf N},t)}{\partial t} =&
\sum_{i,j=1}^m\Bigg(-\frac{\partial}{\partial N_i} \bigg[N_j\alpha_j\mu_{ji} p({\bf N},t)\bigg]
+\frac{1}{2}\frac{\partial^2}{\partial N_i^2} \bigg[N_j\alpha\mu_{ji} p({\bf N},t)\bigg]\Bigg)\\
&+\sum_{i=1}^m\Bigg(\frac{\partial}{\partial N_i} \bigg[N_i\omega_i p({\bf N},t)\bigg]
+\frac{1}{2}\frac{\partial^2}{\partial {N_i}^2} \bigg[N_i\omega_i p({\bf N},t)\bigg]\Bigg).
\end{aligned}
\label{eq:fp-exp}
\end{equation}
Collecting the $\partial /\partial N_i$ and $\partial^2 /\partial^2 N_i$ terms in Eq.~\eqref{eq:fp-exp}, respectively, yields
\begin{align}
A_i({\bf N}) &\coloneq \sum_{j=1}^m N_j\alpha_j\mu_{ji} - N_i\omega_i,\\
\label{eq:ai}
B_i({\bf N}) &\coloneq \sum_{j=1}^m N_j\alpha_j\mu_{ji} + N_i\omega_i,
\end{align}
which can be used to rewrite Eq.~\eqref{eq:fp-exp} in its common form
\begin{equation}
\frac{\partial p({\bf N},t)}{\partial t} =
-\sum_{i=1}^m \frac{\partial}{\partial N_i}\big[A_i({\bf N}) p({\bf N},t)\big]
+\frac{1}{2}\sum_{i=1}^m\frac{\partial^2}{\partial N_i^2}\big[B_{i}({\bf N})p({\bf N},t)\big].
\label{eq:fp-gen}
\end{equation}
Following~\cite{gardiner2009stochastic}, Eq.~\eqref{eq:fp-gen} implies the existence of the It\^{o} SDEs
\begin{equation}
\label{eq:sde-N}
\de N_i = A_i({\bf N})\de t + \sqrt{B_i({\bf N})}\de W_i ,
\end{equation}
where $W_i$ are uncorrelated Wiener processes, \emph{i.e.}, $\mathbb{E}[\de W_i]=0$ and $\mathbb{E}[\de W_i(t)\de W_j(t')]=\delta_{ij}\delta(t-t')\de t$.

We aim at recasting the dynamics in terms of concentrations $n_i\coloneq N_i/N=N_i/\sum_{i=1}^mN_i$.
To do this, we use the multivariate form of It\^{o}'s lemma which, considering that the diffusion matrix is diagonal, yields the following:
\begin{equation}
\de n_i = \left( \sum_{j=1}^m A_j({\bf N})\frac{\partial}{\partial N_j}n_i + \frac{1}{2}\sum_{j=1}^m B_j({\bf N})\frac{\partial^2}{\partial N_j^2}n_i \right)\de t
+ \sum_{j=1}^m \sqrt{B_j({\bf N})}\left(\frac{\partial}{\partial N_j}n_i\right)\de W_j .
\label{eq:ito-multi}
\end{equation}

The first term can be simplified noting that
\begin{equation}
\frac{\partial n_i}{\partial N_j} =
\begin{cases}
\frac{1-n_i}{N} & i=j\\
-\frac{n_i}{N} & i\neq j
\end{cases},
\label{eq:der-cases}
\end{equation}
giving
\begin{equation}
\sum_{j=1}^m A_j({\bf N})\frac{\partial}{\partial N_j}n_i =
A_i({\bf n}) - n_i\sum_{j=1}^m A_j({\bf n}) \eqcolon \Phi_i({\bf n}),
\end{equation}
and, similarly, the second term can be simplified noting that
\begin{equation}
\frac{\partial^2 n_i}{\partial N_j^2} =
\begin{cases}
\frac{2(n_i-1)}{N^2} & i=j\\
\frac{2n_i}{N^2} & i\neq j
\end{cases},
\label{eq:der-cases-2}
\end{equation}
giving
\begin{equation}
\frac{1}{2}\sum_{j=1}^m B_j({\bf N})\frac{\partial^2}{\partial N_j^2}n_i
=-\frac{1}{N}\left(B_i({\bf n}) - n_i\sum_{j=1}^m B_j({\bf n})\right) \eqcolon \Psi_i({\bf n}).
\end{equation}
We note the use of $A_i(\mathbf{n})\coloneq A_i(\mathbf{N})/N$ and $B_i(\mathbf{n})\coloneq B_i(\mathbf{N})/N$ in the above. 
To deal with the third term of Eq.~\eqref{eq:ito-multi} (\textit{i.e.}, the stochastic part), we again use Eq.~\eqref{eq:der-cases} to show that
\begin{equation}
\sqrt{B_j({\bf N})}\frac{\partial n_i}{\partial N_j} = \frac{1}{\sqrt{N}} \times
\begin{cases}
(1-n_i)\sqrt{B_i(\bf n)} & i=j\\
- n_i \sqrt{B_j(\bf n)} & i\neq j
\end{cases}
\eqcolon \frac{1}{\sqrt{N}} q_{ij}({\bf n}).
\end{equation}
Thus Eq.~(\ref{eq:ito-multi}) can equally be expressed as
\begin{align}
    \de n_i &= \left(\Phi_i({\bf n}) + \Psi_i({\bf n})\right)\mathop{\de t} + \frac{1}{\sqrt{N}}\sum_{j=1}^m q_{ij}({\bf n})\mathop{\de W_j},
\end{align}
as appears in the main text. 
\par
To recover an SDE for the total number of individuals, $N$, we can merely sum Eq.~\eqref{eq:sde-N} yielding
\begin{align}
\de N&=\sum_{i=1}^m \de N_i=\sum_{i=1}^m  A_i(\mathbf{N})\de t+\sum_{i=1}^m \sqrt{B_i(\mathbf{N})}\de W_i\nonumber\\
&=N\sum_{i=1}^m n_i(\alpha_i-\omega_i)\de t+\sqrt{N}\sum_{i=1}^m \sqrt{n_i(\alpha_i+\omega_i)}\de W_i\nonumber\\
&=N(\langle\alpha\rangle_{\mathbf{n}}-\langle\omega\rangle_{\mathbf{n}})\de t+\sqrt{N(\langle\alpha\rangle_{\mathbf{n}}+\langle\omega\rangle_{\mathbf{n}}}\de W,
\end{align}
where $\langle\alpha\rangle_{\mathbf{n}}\coloneq\sum_{i=1}^mn_i\alpha_i$ and $\langle\omega\rangle_{\mathbf{n}}\coloneq\sum_{i=1}^mn_i\omega_i$ are the average replication and deterioration rates of the system in configuration $\mathbf{n}$, and where the final expression is in terms of increments of a newly introduced Wiener process $\de W$ which is statistically equivalent to the composite Wiener process $m^{-1/2}\sum_{i=1}^m\de W_i$ and is thus correlated with the $\de W_i$  such that $\mathbb{E}[\de W\de W_i]=m^{-1/2}\de t$.

\section{The Fokker-Planck equation in concentration space}
\label{sec:splitting-fpe}

The SDEs introduced in Appendix~\ref{sec:sse} imply the Fokker-Planck equation
\begin{equation}
\frac{\partial p({\bf n},t)}{\partial t} =
-\sum_{i=1}^m \frac{\partial}{\partial n_i}\left[\left(\Phi_i({\bf n}) + \Psi_i({\bf n})\right) p({\bf n},t)\right]
+\frac{1}{2N}\sum_{i,j=1}^m \frac{\partial^2}{\partial n_i\partial n_j}\big[Q_{ij}({\bf n})p({\bf n},t)\big],
\end{equation}
where $\mathsf{Q}=[Q_{ij}]\coloneq\mathsf{q}\mathsf{q}^T$ and $\mathsf{q}$ is the matrix $[q_{ij}]$.
Our central strategy then relies upon being able to rewrite this as
\begin{align}
\frac{\partial p({\bf n},t)}{\partial t} &=
-\sum_{i=1}^m \frac{\partial}{\partial n_i}\left[\left(\Phi_i({\bf n}) + \Psi_i({\bf n})\right) p({\bf n},t)\right]
+ \frac{1}{2N} \sum_{i,j=1}^m \left[\frac{\partial}{\partial n_i}p({\bf n},t) \frac{\partial}{\partial n_j}Q_{ij}({\bf n})
+ \frac{\partial}{\partial n_i}Q_{ij}({\bf n}) \frac{\partial}{\partial n_j}p({\bf n},t)\right]\nonumber\\
&= -\sum_{i=1}^m \frac{\partial}{\partial n_i}\left[\left(\Phi_i({\bf n}) + \Psi_i({\bf n}) - \frac{1}{2N}\sum_{j=1}^m\frac{\partial}{\partial n_j}Q_{ij}({\bf n}) \right) p({\bf n},t)
-\frac{1}{2N}\sum_{j=1}^m Q_{ij}({\bf n})\frac{\partial}{\partial n_j}p({\bf n},t)\right].
\label{eq:fpe-n}
\end{align}
Eq.~\eqref{eq:fpe-n} is more conveniently written in vector notation as
\begin{equation}
\dot{p}({\bf n}, t) = -\nabla_{\bf n} \cdot {\cal J} , 
\end{equation}
with
\begin{equation}
{\cal J} \coloneq \left( \Phi({\bf n})+\Psi({\bf n})+\hat{\Theta}(\mathbf{n}) \right)p({\bf n}, t) - \frac{1}{2N}\mathsf{Q}({\bf n})\nabla_{\bf n} p({\bf n}, t) ,
\end{equation}
where $\nabla_\mathbf{n}\coloneq(\partial/\partial n_1,\ldots,\partial/\partial n_m)$, $\Phi({\bf n})=(\Phi_1({\bf n}),\ldots,\Phi_m({\bf n}))$,  $\Psi({\bf n})=(\Psi_1({\bf n}),\ldots,\Psi_m({\bf n}))$, and $\hat{\Theta}({\bf n})=(\hat{\Theta}_1({\bf n}),\ldots,\hat{\Theta}_m({\bf n}))$, and where
\begin{align}
\label{eq:theta_i_hat}
\hat{\Theta}_i(\mathbf{n}) =- \frac{1}{2N}\sum_{j=1}^m\frac{\partial}{\partial n_j}Q_{ij}({\bf n}).
\end{align}
It is then natural to attempt to solve for $\dot{p}=0$, assuming absence of stationary flux, implying
\begin{equation}
\frac{1}{2N}\mathsf{Q}({\bf n})\nabla_{\bf n} p({\bf n}, t) = \left( \Phi({\bf n})+\Psi({\bf n})+\hat{\Theta}(\mathbf{n}) \right)p({\bf n}, t),
\end{equation}
and therefore
\begin{equation}
\frac{\nabla_{\bf n} p({\bf n}, t)}{p({\bf n}, t)} = 2N\mathsf{Q}^{-1}({\bf n})\left( \Phi({\bf n})+\Psi({\bf n})+\hat{\Theta}(\mathbf{n}) \right),
\end{equation}
thus associating stationary points in the distribution with zeros of the right-hand side.
\par
However, this reasoning is faulty since it relies on the inversion of the matrix $\mathsf{Q}$, which is singular due to the simplex constraint that $\sum_i n_i=1$. 
To proceed, as per the main text, we exploit the property that solution is symmetrically singular in the sense that $p(n_1,..., n_m)=p_{\backslash i}(n_1,...,n_{i-1},n_{i+1},...,n_m)\delta(n_i-(1-\sum_{k\neq i}n_k))$, for all $i$. As such we can eliminate an arbitrary coordinate (we choose $m$ without loss of generality) producing a FPE for $\tilde{p}(\tilde{\mathbf{n}})\equiv p_{\backslash m}(\{n_1, \dots , n_{m-1}\})$. This distribution is governed by a FP operator exactly as the above, but with all instances of $n_m$ replaced as $n_m\to 1-\sum_{k=1}^{m-1}n_k$ such that all quantities become functions of $\tilde{\mathbf{n}}=\{n_1,\ldots,n_{m-1}\}$, and with the appropriate $m$-th element of the vectors, and $m$-th row/column of the matrices, removed. We write the resulting reduced matrix as $\mathcal{Q}(\tilde{\mathbf{n}})=[\mathsf{Q}(\mathbf{n})]_{1:m-1,1:m-1}$, comprised of elements $\mathcal{Q}_{ij}$. For vanishing stationary flux the above reasoning then becomes valid and we identify
\begin{equation}
   \frac{\nabla_{\tilde{\bf n}} \tilde{p}(\tilde{\bf n}, t)}{\tilde{p}(\tilde{\bf n}, t)} = 2N\mathcal{Q}^{-1}(\tilde{\bf n})\left( \Phi(\tilde{\bf n})+\Psi(\tilde{\bf n}) +\Theta(\tilde{\bf n})\right),
\end{equation}
but now with $\Phi(\tilde{\bf n})=(\Phi_1(\tilde{\bf n}),\ldots,\Phi_{m-1}(\tilde{\bf n}))$, \emph{etc.}, and with the noise induced term  now given by
\begin{align}
    \label{eq:theta_i}
   \Theta_i(\tilde{\mathbf{n}})&\coloneq - \frac{1}{2N}\sum_{j=1}^{m-1}\frac{\partial}{\partial n_j}\mathcal{Q}_{ij}(\tilde{\bf n}),
\end{align}
constructed from the reduced matrix $\mathcal{Q}(\tilde{\mathbf{n}})$, and with its constituent sum ranging from $1$ to $m-1$. This distinction is crucial and we recognise that $\Theta(\tilde{\mathbf{n}})\neq \hat{\Theta}(\mathbf{n})$ (\emph{I.e.} Eqs.~(\ref{eq:theta_i}) and (\ref{eq:theta_i_hat}) are not equal).  
This then allows us to identify the total effective force
\begin{align}
   F(\tilde{\mathbf{n}})&\coloneq \Phi(\tilde{\bf n})+\Psi(\tilde{\bf n}) +\Theta(\tilde{\bf n})
\end{align}
which vanishes at stationary points of the limiting distribution in the case of zero stationary current, and is proportional to the total non-dispersive flux in the general case. 

\subsection*{Example solution, equivalent variants, ${\bm m}\ {\bf =2}$}
It is instructive to consider a simple, exactly solvable, example wherein we can illustrate the important qualitative features that the noise-induced forces can exert on the system. In particular, we aim to show: 1) that knowing just the solution at the level of the mean is not sufficient to describe the overall behaviour exhibited by the system, and 2) that at low population size $N$ and/or high fidelity $\hat{\mu}\coloneq\mu_{ii}$, the probability distribution undergoes a qualitative change we associate with the fidelity catastrophe.
\par
\begin{figure}[!t]
   \centering
   \includegraphics[width=0.85\textwidth]{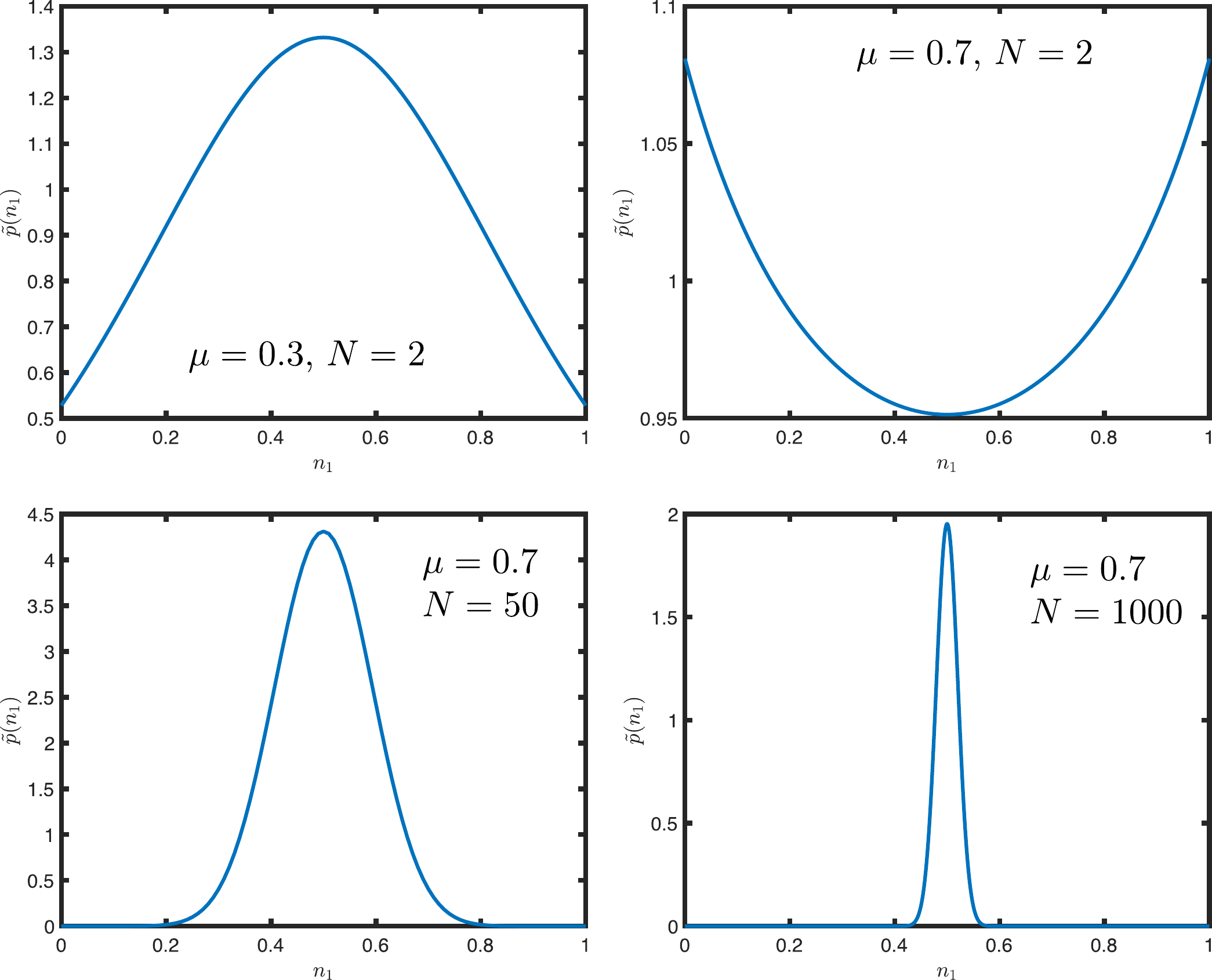}
   \caption{Exact probability density functions for the $m=2$ equivalent variant system for $\alpha=\omega=1$. The behaviour of the system depends strongly on $N$ and $\hat{\mu}$, but all solutions share the mean value of $\langle n_1\rangle=1/2$.}
   \label{fig:exact_2d}
\end{figure}

To this end, let us consider a simple example of equivalent species (equal replication and deterioration rates) in the case of just two variants, $m=2$. In this case we can enumerate all required objects as
\begin{subequations}
\begin{align}
\mu_{11}&=\mu_{22}=\hat{\mu},\\
\mu_{12}&=\mu_{21}=1-\hat{\mu},\\
A_1(\mathbf{n})&= \alpha\hat{\mu} n_1 + \alpha (1-\hat{\mu})n_2 - n_1\omega,\\
A_2(\mathbf{n})&= \alpha\hat{\mu} n_2 + \alpha (1-\hat{\mu})n_1 - n_2\omega,\\
B_1(\mathbf{n})&= \alpha\hat{\mu} n_1 + \alpha (1-\hat{\mu})n_2 + n_1\omega,\\
B_2(\mathbf{n})&= \alpha\hat{\mu} n_2 + \alpha (1-\hat{\mu})n_1 + n_2\omega,\\
\Phi_1(\mathbf{n})&=\alpha\hat{\mu} n_1 + \alpha (1-\hat{\mu})n_2 - n_1\alpha,\\
\Psi_1(\mathbf{n})&=-\Phi_1(\mathbf{n})/N,\\
\Phi_2(\mathbf{n})&=\alpha\hat{\mu} n_2 + \alpha (1-\hat{\mu})n_1 - n_2\alpha,\\
\Psi_2(\mathbf{n})&=-\Phi_2(\mathbf{n})/N,\\
q_{11}(\mathbf{n})&=(1-n_1)\sqrt{B_1},\\
q_{12}(\mathbf{n})&=-n_1\sqrt{B_2},\\
q_{12}(\mathbf{n})&=-n_2\sqrt{B_1},\\
q_{22}(\mathbf{n})&=(1-n_2)\sqrt{B_2},\\
Q_{ij}(\mathbf{n})&=\sum_{k=1}^2q_{ik}(\mathbf{n})q_{jk}(\mathbf{n}).
\end{align}
\end{subequations}
As per Sec.~\ref{sec:splitting-fpe}, constructing the Fokker-Planck equation in terms of $\mathbf{n}=\{n_1,n_2\}$ is problematic owing to the constraint $n_1+n_2=1$ leading to a singular diffusion matrix, $\mathsf{Q}$. To solve for this system we need to consider the underlying non-singular sub-system, which we can do by eliminating $n_2$ by writing $n_2\to 1-n_1$, such that we deal with $\tilde{\mathbf{n}}=n_1$. This then also requires the consideration of the sub-matrix, $\mathcal{Q}(n_1)$, which is now simply the single element, $\mathcal{Q}(n_1)=\mathcal{Q}_{11}(n_1)=Q_{11}(n_1,1-n_1)$. As such we have the following 1D system (for which the limiting state must have vanishing flux):
\begin{align}
\mathcal{Q}(n_1)&=-\left(n_1-1\right) n_1 \omega -\alpha  \left(\hat{\mu} +(4\hat{\mu} -3) \left(n_1-1\right) n_1-1\right),\\
F_1(n_1)&=\Phi_1(n_1)+\Psi_1(n_1)+\Theta_1(n_1)\nonumber\\
&=\Phi_1(n_1)+\Psi_1(n_1)-(2N)^{-1}\partial_{n_1}\mathcal{Q}(n_1)\nonumber\\
&=\frac{\left(2 n_1-1\right) (\alpha  (2\hat{\mu} +2 (\hat{\mu} -1) N-1)+\omega )}{2 N},\\
\frac{\partial_{n_1}\tilde{p}(n_1)}{\tilde{p}(n_1)}&=2N\frac{F_1(n_1)}{\mathcal{Q}(n_1)}\nonumber\\
&=-\frac{\left(2 n_1-1\right) (\alpha  (2\hat{\mu} +2 (\hat{\mu} -1) N-1)+\omega )}{\left(\alpha  \left(\hat{\mu} +(4\hat{\mu}
   -3) \left(n_1-1\right) n_1-1\right)+\left(n_1-1\right) n_1 \omega \right)},
\end{align}
with solution, up to a normalising constant,
\begin{align}
\tilde{p}(n_1)&\propto\exp\left[\int dn_1\;2N\frac{F_1(n_1)}{\mathcal{Q}(n_1)}\right]
=\left(\alpha  (\hat{\mu} -1)+\left(n_1-1\right) n_1 (\alpha  (4\hat{\mu} -3)+\omega )\right){}^{-\frac{\alpha  (2\hat{\mu} +2 (\hat{\mu} -1)
  N-1)+\omega }{(\alpha  (4\hat{\mu} -3)+\omega )}}.
\end{align}
On the level of the mean, this distribution has identical behaviour for all parameters, with the mean simply equal to $\langle n_1\rangle=1/2$, by symmetry. However, it exhibits very different behaviour depending on the choice of parameters. These are illustrated in Fig.~\ref{fig:exact_2d} for the case $\alpha=\omega=1$. Note, however, that as $N\to\infty$ the distribution tends to an increasingly narrow Gaussian centred on $n_1=1/2$, as expected from the deterministic version of the Eigen model. 
\par
At low $N$, however, we particularly highlight that for various distinct parameter values the probability can be either concentrated on the mean, \emph{i.e.}, around $n_1=n_2=1/2$, or bimodally, far from the mean, in the regions corresponding to homogeneous, one species dominance, \emph{i.e.}, $n_1=1$, $n_2=0$, and $n_1=0$, $n_2=1$. Specifically, for small $N$, where we expect the noise-induced behaviour to dominate, the value of $\hat{\mu}$ controls the shape of the distribution between these two extremes. Where the distribution is bimodal in this way --- at low $N$ and high $\hat{\mu}$ --- we say that the system has suffered from the fidelity catastrophe. 
\par
This simple case is presented since it is exactly solvable, however we wish to understand systems with much higher dimensionality. To do so we implement the methodology expressed in the main text, whereby, in lieu of full solutions of the Fokker-Planck equation, we seek effective forces which can be associated with the non-dispersive component of the limiting flux, even if we cannot state the full distribution. The expression for these forces, first in general, and then along illustrative lines in phase space, are the subject of the following sections.

\section{Forces in the simplex}
As per the main text, we have identified a total effective force vector, $F(\tilde{\mathbf{n}})$, comprised of $m$ components, $F_i(\tilde{\mathbf{n}})$, $i\in\{1,\ldots,m\}$, one for each variant (We note, only $m-1$ are required in the FPE due to the simplex constraint). Each such component, in turn, is formed from three contributions
\begin{align}
   F_i(\tilde{\mathbf{n}})&\coloneq \Phi_i(\tilde{\mathbf{n}})+\Psi_i(\tilde{\mathbf{n}})+\Theta_i(\tilde{\mathbf{n}}),
\end{align}
with the latter contribution identified as originating from the multiplicative nature of the noise, derived from the reduced diffusion matrix.
\par
Here we aim to derive expressions for these forces at any point within the simplex, making only the assumptions $\mu_{ii}=\mu_{jj}$ and $\mu_{ij}=\mu_{ji}$ for all $i,j\in\{1,\ldots,m\}$. We then note the essential corollary $\sum_{j=1}^m\mu_{ij}=\sum_{j=1}^m\mu_{ji}=1$.
\par
To do so, and for the sake of readability, we first define the following quantities:
\begin{equation}
\bar{\alpha}\coloneq\frac{1}{m}\sum_{j=1}^m\alpha_j \ \ \ \ \textrm{and}\ \ \ \  \langle\alpha\rangle_{\mathbf{n}}\coloneq\sum_{j=1}^mn_j\alpha_j \ ,
\end{equation}
\begin{equation}
\bar{\omega}\coloneq\frac{1}{m}\sum_{j=1}^m\omega_j \ \ \ \ \textrm{and}\ \ \ \  \langle\omega\rangle_{\mathbf{n}}\coloneq\sum_{j=1}^mn_j\omega_j \ ,
\end{equation}
\begin{equation}
\mathcal{T}_{ji}\coloneq\alpha_j\mu_{ji} \ \ \ \ \textrm{and}\ \ \ \  \langle\mathcal{T}_{\cdot a}\rangle_{\mathbf{n}}\coloneq\sum_{j=1}^mn_j\mathcal{T}_{ja}=\sum_{j=1}^m\alpha_jn_j\mu_{ja}=B_a(\mathbf{n})-n_a\omega_a \ ,
\end{equation}
\begin{equation}
\gamma_i\coloneq\sum_{j=1}^m\mathcal{T}_{ji} \ \ \ \ \textrm{and}\ \ \ \  \bar{\gamma}\coloneq\frac{1}{m}\sum_{j=1}^m\gamma_j =\frac{1}{m}\sum_{j=1}^m\sum_{k=1}^m\alpha_k\mu_{kj}=\bar{\alpha} \ .
\end{equation}
We also recognise that
\begin{align}
\sum_{i=1}^mA_i(\mathbf{n})&=\sum_{i=1}^m\left(\sum_{j=1}^m\alpha_jn_j\mu_{ji}-n_i\omega_i\right)=\langle\alpha\rangle_{\mathbf{n}}-\langle\omega\rangle_{\mathbf{n}},\\
\sum_{i=1}^mB_i(\mathbf{n})&=\sum_{i=1}^m\left(\sum_{j=1}^m\alpha_jn_j\mu_{ji}+n_i\omega_i\right)=\langle\alpha\rangle_{\mathbf{n}}+\langle\omega\rangle_{\mathbf{n}},
\end{align}
and
\begin{equation}
\sum_{i=1}^m{\gamma}_i=\sum_{i=1}^m\sum_{j=1}^m\alpha_j\mu_{ji}=\sum_{j=1}^m\alpha_j =m\bar{\alpha}.
\end{equation}
Recalling $\Phi_i({\bf n}) \coloneq A_i({\bf n}) - n_i\sum_{j=1}^m A_j({\bf n})$, with $A_i({\bf n})\coloneq\sum_{j=1}^m n_j\alpha_j\mu_{ji} - n_i\omega_i$, and $\Psi_i({\bf n}) \coloneq -N^{-1}(B_i({\bf n}) - n_i\sum_{j=1}^m B_j({\bf n}))$, with $B_i({\bf n})\coloneq\sum_{j=1}^m n_j\alpha_j\mu_{ji} + n_i\omega_i$, we can immediately derive expressions for the two contributions deriving from the deterministic component of the dynamics
\begin{align}
   \Phi_i(\tilde{\mathbf{n}})&\equiv\Phi_i(\mathbf{n})\nonumber\\
   &=A_i-n_i\left(\langle\alpha\rangle_\mathbf{n}-\langle\omega\rangle_\mathbf{n}\right)\nonumber\\
   &=\langle\mathcal{T}_{\cdot i}\rangle_{\mathbf{n}}-n_i\left(\langle\alpha\rangle_\mathbf{n}+(\omega_i-\langle\omega\rangle_\mathbf{n})\right)\\
   \Psi_i(\tilde{\mathbf{n}})&\equiv\Psi_i(\mathbf{n}),\nonumber\\
   &=-N^{-1}B_i+N^{-1}n_i\left(\langle\alpha\rangle_\mathbf{n}+\langle\omega\rangle_\mathbf{n}\right)\nonumber\\
   &=-N^{-1}\langle\mathcal{T}_{\cdot i}\rangle_{\mathbf{n}}+N^{-1}n_i\left(\langle\alpha\rangle_\mathbf{n}-(\omega_i-\langle\omega\rangle_\mathbf{n})\right),
\end{align}
since these terms are invariant under either $\mathbf{n}$ or $\tilde{\mathbf{n}}$ (subject to the substitution $n_m\to 1-\sum_{i=1}^{m-1}n_i$, assuming removal of the $m$th component).
\par
Calculating the noise-induced component, however, is slightly more involved, since it is defined in terms of derivatives of the reduced matrix $\mathcal{Q}$. Explicitly, the form of this force component (assuming the reduction in state space by removing $n_m$) is given by
\begin{equation}
\Theta_i(\tilde{\mathbf{n}}) \coloneq -\frac{1}{2N}\sum_{j=1}^{m-1}\partial_{n_j} \mathcal{Q}_{ij}(\tilde{\mathbf{n}}).
\label{eq:thetai}
\end{equation}
To start, we derive the expressions for $Q_{ii}(\mathbf{n})$ and $Q_{ij}(\mathbf{n})$, noting that these are in terms of the full matrix, $\mathsf{Q}$, and vector, $\mathbf{n}$:
\begin{equation}
{Q}_{ii}(\mathbf{n})=\sum_{j=1}^mq_{ij}(\mathbf{n})q_{ij}(\mathbf{n})=(1-n_i)^2B_i(\mathbf{n})+n_i^2\sum_{j=1,j\neq i}^m B_j(\mathbf{n}) ,
\end{equation}
\begin{equation}
{Q}_{ij}(\mathbf{n})=\sum_{k=1}^mq_{ik}(\mathbf{n})q_{jk}(\mathbf{n})=-(1-n_i)n_jB_i(\mathbf{n})-(1-n_j)n_iB_j(\mathbf{n})+n_in_j\sum_{k=1,k\notin\{i,j\}}^m B_k(\mathbf{n}).
\end{equation}
Next, we consider the $Q_{ij}(\mathbf{n})$ as part of the reduced matrix, $\mathcal{Q}_{ij}(\mathbf{\tilde{n}})$, by replacing all instances of $n_m$ as per
\begin{equation}
n_m\to 1-\sum_{j=1}^{m-1}n_j.
\end{equation}
This change manifests itself in the forms of the derivatives that appear in Eq.~\eqref{eq:thetai}, noting that the use of the above substitution leads to special cases where the $m$-th coordinate is involved such that 
\begin{align}
   \partial_{n_i} n_i&=1,\,\quad\forall\,i\in\{1,\ldots,m-1\}\\
\partial_{n_i} n_m&=-1,\,\quad\forall\,i\in\{1,\ldots,m-1\}
\end{align}
and 
\begin{align}
   \partial_{n_i} B_i(\mathbf{n})&=\alpha_i\mu_{ii}+\omega_i-\alpha_m \mu_{mi},\\
   \partial_{n_j} B_i(\mathbf{n}) &= 
   \begin{cases}
       \alpha_j \mu_{ji}-\alpha_i\mu_{ii}-\omega_{i} & i=m,\,j\neq i,\\
       \alpha_j\mu_{ji}-\alpha_m \mu_{mi} & i\neq m,\,j\neq i.
   \end{cases}
\end{align}
We can then use these to consider the derivatives of the $\mathcal{Q}_{ij}(\tilde{\mathbf{n}})$, noting that the constituent sums for each element of $\mathcal{Q}$ are still over $\{1,\ldots,m\}$ since the individual $\mathcal{Q}_{ij}$ terms are still formed from $m$ noise terms.
\par
First we consider $\partial_i{\mathcal{Q}}_{ii}(\mathbf{\tilde{n}})$, noting that the $i=m$ case is not needed since it does not appear in Eq.~\eqref{eq:thetai}:
\begin{align}
\partial_{n_i}\mathcal{Q}_{ii}(\mathbf{\tilde{n}})&=\partial_{n_i}\left[(1-n_i)^2B_i(\mathbf{n})+n_i^2\sum_{j=1,j\neq i}^mB_j(\mathbf{n})\right]\nonumber\\
&=\partial_{n_i}\left[(1-n_i)^2B_i(\mathbf{n})+n_i^2B_m(\mathbf{n})+n_i^2\sum_{j=1,j\notin \{i,m\}}^m B_j(\mathbf{n})\right]\nonumber\\
&=-2(1-n_i)B_i(\mathbf{n})+(1-n_i)^2\partial_{n_i} B_i(\mathbf{n})+2n_iB_m(\mathbf{n})+n_i^2\partial_{n_i}B_m(\mathbf{n})+2n_i\sum_{j=1,j\notin \{i,m\}}^mB_j(\mathbf{n})\nonumber\\
&\qquad+n_i^2\sum_{j\notin \{i,m\}}\partial_iB_j(\mathbf{n})\nonumber\\
&=-2(1-n_i)B_i(\mathbf{n})+(1-n_i)^2\left(\alpha_i\mu_{ii}+\omega_i-\alpha_m \mu_{mi}\right)+n_i^2\left(\alpha_i \mu_{im}-\alpha_m\mu_{mm}-\omega_{m}\right)\nonumber\\
&\qquad+2n_i\sum_{j=1,j\neq i}^mB_j(\mathbf{n})+n_i^2\sum_{j\notin \{i,m\}}\left(\alpha_i\mu_{ij}-\alpha_m \mu_{mj}\right)\nonumber\\
&=-2 B_i(\mathbf{n})+2 n_i \left(\langle\alpha\rangle_{\mathbf{n}}+\langle\omega\rangle_{\mathbf{n}}-\alpha_i \mu_{{ii}}-\omega_i+\alpha_m \mu _{mi}\right)+\alpha _i \mu _{{ii}}+n_i^2 \left(\alpha _i+\omega_i-\alpha_m-\omega_m\right)\nonumber\\
&\qquad+\omega_i-\alpha_m \mu_{mi}.
\end{align}

Next we consider $\partial_{n_j}\mathcal{Q}_{ij}(\mathbf{\tilde{n}})$, noting that neither the $i=m$ nor $j=m$ cases are needed, but the constituent sum over $k$ does include the $m$-th term:
\begin{align}
\partial_{n_j}\mathcal{Q}_{ij}(\mathbf{\tilde{n}})&=\partial_{n_j}\left[-(1-n_i)n_jB_i(\mathbf{n})-(1-n_j)n_iB_j(\mathbf{n})+n_in_j\sum_{k=1,k\notin\{ i,j\}}^mB_k(\mathbf{n})\right]\nonumber\\
&=-(1-n_i)B_i(\mathbf{n})-(1-n_i)n_j\partial_{n_j}B_i(\mathbf{n})+n_iB_j(\mathbf{n})-(1-n_j)n_i\partial_{n_j}B_j(\mathbf{n})+n_i\sum_{k=1,k\notin\{ i,j\}}^mB_k(\mathbf{n})\nonumber\\
&\qquad+n_in_j\partial_{n_j}B_m(\mathbf{n})+n_in_j\sum_{k=1,k\notin\{ i,j,m\}}^m\partial_{n_j}B_k(\mathbf{n})\nonumber\\
&=-(1-n_i)B_i(\mathbf{n})-(1-n_i)n_j\left(\alpha_j\mu_{ji}-\alpha_m \mu_{mi}\right)+n_iB_j(\mathbf{n})-(1-n_j)n_i\left(\alpha_j\mu_{jj}+\omega_i-\alpha_m \mu_{mj}\right)\nonumber\\
&\qquad+n_i(\langle\alpha\rangle_{\mathbf{n}}+\langle\omega\rangle_{\mathbf{n}}-B_i(\mathbf{n})-B_j(\mathbf{n}))+n_in_j\left(\alpha_j \mu_{jm}-\alpha_m\mu_{mm}-\omega_{m}\right)\nonumber\\
&\qquad+n_in_j\sum_{k=1,k\notin\{ i,j,m\}}^m\left(\alpha_j\mu_{jk}-\alpha_m \mu_{mk}\right)\nonumber\\
&=-B_i(\mathbf{n})+\alpha _j \left(n_i \left(n_j-\mu_{{ii}}\right)-n_j \mu_{{ji}}\right)+\alpha_m \left(n_i \left(\mu_{{mj}}-n_j\right)+n_j \mu_{{mi}}\right)\nonumber\\
&\qquad+n_i \left(-n_j \omega_m+\left(n_j-1\right) \omega_j+\langle\alpha\rangle_{\mathbf{n}}+\langle\omega\rangle_{\mathbf{n}}\right).
\end{align}
We now consider the off-diagonal component of the sum in Eq.~\eqref{eq:thetai}, noting this excludes $m$:
\begin{align}
\sum_{j=1,j\notin\{i,m\}}^m\partial_{n_j}\mathcal{Q}_{ij}(\mathbf{\tilde{n}})&=\sum_{j=1,j\notin\{i,m\}}^m[-B_i(\mathbf{n})+\alpha _j \left(n_i \left(n_j-\mu _{{ii}}\right)-n_j \mu_{{ji}}\right)+\alpha_m \left(n_i \left(\mu_{{mj}}-n_j\right)+n_j \mu_{{mi}}\right)\nonumber\\
&\qquad+n_i \left(-n_j \omega_m+\left(n_j-1\right) \omega_j+\langle\alpha\rangle_{\mathbf{n}}+\langle\omega\rangle_{\mathbf{n}}\right)]\\
&=-(m-2)B_i(\mathbf{n})+n_i(\langle\alpha\rangle_{\mathbf{n}}-n_i\alpha_i-n_m\alpha_m)-n_i\mu_{ii}(m\bar{\alpha}-\alpha_i-\alpha_m)-\sum_{j=1,j\notin\{i,m\}}^m\alpha_jn_j\mu_{ji}\nonumber\\
&\qquad+\alpha_mn_i(1-\mu_{mi}-\mu_{mm})-\alpha_mn_i(1-n_i-n_m)+\alpha_m\mu_{mi}(1-n_i-n_m)\nonumber\\
&\qquad-\omega_mn_i(1-n_i-n_m)+n_i(\langle\omega\rangle_{\mathbf{n}}-n_i\omega_i-n_m\omega_m)-n_i(m\bar{\omega}-\omega_i-\omega_m)\nonumber\\
&\qquad+(m-2)n_i(\langle\alpha\rangle_{\mathbf{n}}+\langle\omega\rangle_{\mathbf{n}})\\
&=n_i \left(\mu_{{ii}} \left(2 \alpha_i-m\bar{\alpha} \right)+(m-1) (\langle\alpha\rangle_{\mathbf{n}}+\langle\omega\rangle_{\mathbf{n}})+2\omega _i-2 \alpha_m \mu_{mi}-m \bar{\omega}\right)-(m-1) B_i(\mathbf{n})\nonumber\\
&\qquad+n_i^2 \left(-\alpha_i-\omega_i+\alpha_m+\omega_m\right)+\alpha_m \mu_{{mi}}.
\end{align}
Finally, we add the $\partial_i\mathcal{Q}_{ii}(\mathbf{\tilde{n}})$ component to get
\begin{align}
-2N\Theta_i(\tilde{\mathbf{n}}) &= \sum_{j=1}^{m-1}\partial_{n_j} \mathcal{ Q}_{ij}(\tilde{\mathbf{n}})\nonumber\\
&=-(m+1)B_i(\mathbf{n})+(m+1)n_i(\langle\alpha\rangle_{\mathbf{n}}+\langle\omega\rangle_{\mathbf{n}})+\mu_{ii}(\alpha_i-n_im\bar{\alpha})+(\omega_i-n_im\bar{\omega})\nonumber\\
&=-(m+1)(\langle\mathcal{T}_{\cdot i}\rangle_{\mathbf{n}}+n_i\omega_i)+(m+1)n_i(\langle\alpha\rangle_{\mathbf{n}}+\langle\omega\rangle_{\mathbf{n}})+\mu_{ii}(\alpha_i-n_im\bar{\alpha})+(\omega_i-n_im\bar{\omega})
\end{align}
which is independent of any parameter or variable to do with the eliminated coordinate, but defines the noise-induced force in any Cartesian component of the system (\emph{i.e.}, running $1$ to $m$).
\par
Finally, we can now express all components of the total force, which can be evaluated at any point within the simplex:
\begin{empheq}[box=\fbox]{align}
   F_i(\tilde{\mathbf{n}})&=\Phi_i(\tilde{\mathbf{n}})+\Psi_i(\tilde{\mathbf{n}})+\Theta_i(\tilde{\mathbf{n}}),\\
   \Phi_i(\tilde{\mathbf{n}})&=\langle\mathcal{T}_{\cdot i}\rangle_{\mathbf{n}}-n_i\left(\langle\alpha\rangle_\mathbf{n}+(\omega_i-\langle\omega\rangle_\mathbf{n})\right),\\
   \Psi_i(\tilde{\mathbf{n}})&=-N^{-1}\left[\langle\mathcal{T}_{\cdot i}\rangle_{\mathbf{n}}-n_i\left(\langle\alpha\rangle_\mathbf{n}-(\omega_i-\langle\omega\rangle_\mathbf{n})\right)\right],\\
   \Theta_i(\tilde{\mathbf{n}})&=(2N)^{-1}\left[(m+1)(\langle\mathcal{T}_{\cdot i}\rangle_{\mathbf{n}}+n_i\omega_i)-(m+1)n_i(\langle\alpha\rangle_{\mathbf{n}}+\langle\omega\rangle_{\mathbf{n}})-\mu_{ii}(\alpha_i-n_im\bar{\alpha})-(\omega_i-n_im\bar{\omega})\right].
\end{empheq}
We stress, despite its expression as a function of $\tilde{\mathbf{n}}\coloneq\{n_1,\ldots,n_{m-1}\}$, there are still $m$ component terms to each force --- but they are constrained such that they lie within the $m-1$ dimensional simplex, such that only $m-1$ terms are required to uniquely determine it.

\section{Forces along the bisecting simplex line}

Here we demonstrate how to evaluate the forces in a particular direction within the simplex, and in particular to evaluate it along a line that bisects the centre of the simplex and one of its vertices.
Let $c$ be the coordinate we are investigating (\textit{i.e.}, the vertex of the simplex), and $b=\{1,\ldots,m\}\setminus \{c\}\equiv \{1,\ldots,c-1,c+1,\ldots,m\}$ the set of all other coordinates such that we define the parametric line, $\mathbf{n}^{(c)}(x)$, as
\begin{equation}
n_i^{(c)}(x)\coloneq\begin{cases}
n_c=x & i=c,\\
n_b=\frac{1-x}{m-1} & i\in b.
\end{cases}
\end{equation}
We use this to find simplified parametrised forms of key variables, starting with
\begin{align}
B_i(x)&=\alpha_c n_c(x)\mu_{ci}+\sum_{j\neq c}\alpha_j n_b(x)\mu_{ji}+n_i(x)\omega_i\\
&=\alpha_cn_{c}(x)\mu_{ci}+n_b(x)({\gamma}_i-\alpha_c\mu_{ci})+n_i(x)\omega\\
&=\alpha_c\mu_{ci}(n_c(x)-n_b(x))+n_b(x){\gamma}_i+n_i(x)\omega\\
&=\alpha_c\mu_{ci}\frac{mx-1}{m-1}+\frac{1-x}{m-1}{\gamma}_i+n_i(x)\omega\\
&=\begin{cases}
\frac{\alpha_c\mu_{cc}(mx-1)+{\gamma}_c(1-x)}{m-1}+x\omega_c&i=c,\\
\frac{\alpha_i\mu_{ci}(mx-1)+({\gamma}_i+\omega_i)(1-x)}{m-1} &i\in b.
\end{cases}
\end{align}
We also consider the parametrised form of the instantaneous averages of the replication and deterioration rates
\begin{align}
\langle\alpha\rangle_{\mathbf{n}}&=x\alpha_c+\frac{1-x}{m-1}(m\bar{\alpha}-\alpha_c),\\
\langle\omega\rangle_{\mathbf{n}}&=x\omega_c+\frac{1-x}{m-1}(m\bar{\omega}-\omega_c).
\end{align}

This allows us to evaluate the functions on the line in terms of the parametrisation variable $x$, but we also need to project them along this line.
We do this by taking the inner product with the direction vector, $\mathbf{u}^{(c)}$, into the $c$ vertex of the simplex:
\begin{equation}
u_i^{(c)}\coloneq\begin{cases}
u_c^{(c)}=\sqrt{\frac{m-1}{m}} & i=c,\\
u_b^{(c)}=-\sqrt{\frac{m-1}{m}}\frac{1}{m-1} & i\in b.
\end{cases}
\end{equation}
Using such a construction, we first compute the noise-induced force in the $\mathbf{u}^{(c)}$ direction along the line as
{
   \allowdisplaybreaks
\begin{align}
f^{(c)}_\Theta(x)&\coloneq\Theta_c(\mathbf{n}^{(c)}(x))u^{(c)}_c+\sum_{i\in b}\Theta_i(\mathbf{n}^{(c)}(x))u^{(c)}_b\nonumber\\
&=\sqrt{\frac{m-1}{4mN^2}}\Bigg[(m+1)B_c(x)-\frac{m+1}{m-1}\sum_{i\in b}B_b(x)\nonumber\\
&\qquad-(m+1)(\langle\alpha\rangle_{\mathbf{n}}+\langle\omega\rangle_{\mathbf{n}})\left(n_c(x)-\frac{1}{m-1}\sum_{i\in b}n_b(x)\right)\nonumber\\
&\qquad+(\mu_{cc}m\bar{\alpha}+m\bar{\omega})\left(n_c(x)-\frac{1}{m-1}\sum_{j\in b}n_b(t)\right)\nonumber\\
&\qquad-(\mu_{cc}\alpha_c+\omega_c)+\frac{1}{m-1}\sum_{i\in b}(\mu_{cc}\alpha_i+\omega_i)\Bigg]\nonumber\\
&=\sqrt{\frac{m-1}{4mN^2}}\Bigg[(m+1)\left(\frac{\alpha_c\mu_{cc}(mx-1)+{\gamma}_c(1-x)}{m-1}+t\omega_c\right)\nonumber\\
&\qquad-\frac{m+1}{m-1}\sum_{i\in b}\left(\frac{\alpha_i\mu_{ci}(mx-1)+({\gamma}_i+\omega_i)(1-x)}{m-1} \right)\nonumber\\
&\qquad-(m+1)(\langle\alpha\rangle_{\mathbf{n}}+\langle\omega\rangle_{\mathbf{n}})\left(\frac{mx-1}{m-1}\right)+(\mu_{cc}m\bar{\alpha}+m\bar{\omega})\left(\frac{mx-1}{m-1}\right)\nonumber\\
&\qquad-(\mu_{cc}\alpha_c+\omega_c)+\frac{1}{m-1}(\mu_{cc}m\bar{\alpha}-\mu_{cc}\alpha_c+m\bar{\omega}-\omega_c)\Bigg]\nonumber\\
&=\sqrt{\frac{m-1}{4mN^2}}\Bigg[(m+1)\left(\frac{\alpha_c\mu_{cc}(mx-1)+{\gamma}_c(1-x)}{m-1}+t\omega_c\right)\nonumber\\
&\qquad-\frac{m+1}{m-1}\left(\frac{\alpha_i(1-\mu_{cc})(mx-1)+(m\bar{\alpha}-{\gamma}_c+m\bar{\omega}-\omega_c)(1-x)}{m-1} \right)\nonumber\\
&\qquad-(m+1)(\langle\alpha\rangle_{\mathbf{n}}+\langle\omega\rangle_{\mathbf{n}})\left(\frac{mx-1}{m-1}\right)+(\mu_{cc}m\bar{\alpha}+m\bar{\omega})\left(\frac{mx-1}{m-1}\right)\nonumber\\
&\qquad-(\mu_{cc}\alpha_c+\omega_c)+\frac{1}{m-1}(\mu_{cc}m\bar{\alpha}-\mu_{cc}\alpha_c+m\bar{\omega}-\omega_c)\Bigg]\nonumber\\
&=\sqrt{\frac{m}{4N^2(m-1)^3}}\bigg[\alpha_c\left(m\mu_{cc}(mx+x-2)+(m+1)x(1-mx)\right)-\omega_c((m+1)m(x-1)x+m-1)\nonumber\\
&\qquad+\bar{\alpha}(m-1)mx\mu_{cc}+(m+1)(x-1)(\bar{\alpha}mx-{\gamma}_c)+mx\bar{\omega}(mx+x-2)\bigg].
\label{eq:noise-force-line}
\end{align}
}
\par
Similarly, we can compute the other force terms. We first compute the requisite terms in terms of the parameter $x$
\begin{align}
A_i(x)&=\alpha_c n_{c}(x)\mu_{ci}+\sum_{j\neq a}\alpha_jn_b(x)\mu_{ji}-n_i(x)\omega_i\nonumber\\
&=\begin{cases}
\frac{{\gamma}_c(1-x)+\alpha_c\mu_{cc}(mx-1)}{m-1} -t\omega_c&i=c\\
\frac{({\gamma}_i-\omega_c)(1-x)+\alpha_c\mu_{ci}(mx-1)}{m-1} &i\in b.
\end{cases}\\
\Phi_i(x)&=A_i(x)-n_i(x)(\langle\alpha\rangle_{\mathbf{n}}(x)-\langle\omega\rangle_{\mathbf{n}}(x))\nonumber\\
&=\begin{cases}
\frac{{\gamma}_c(1-x)+\alpha_c\mu_{cc}(mx-1)}{m-1} -x\omega_c-x(\langle\alpha\rangle_{\mathbf{n}}(x)-\langle\omega\rangle_{\mathbf{n}}(x))&i=c\\
\frac{({\gamma}_i-\omega_c)(1-x)+\alpha_c\mu_{ci}(mx-1)}{m-1} -\frac{1-x}{m-1}(\langle\alpha\rangle_{\mathbf{n}}(x)-\langle\omega\rangle_{\mathbf{n}}(x))&i\in b.
\end{cases}\nonumber\\
&=\begin{cases}
\frac{m (x-1) x \bar{\alpha }-m (x-1) t \bar{\omega }-\alpha _c \mu _{\text{cc}}+mx
   \alpha _c \mu _{cc}-m x^2 \alpha _c + m x^2 \omega _c-mx \omega _c+x \alpha
   _c-(x-1) {\gamma}_c}{m-1}&i=c\\
      -\frac{(x-1) \left(m (x-1) \bar{\alpha }+\bar{\omega } (m-mx)-(mx-1) \left(\alpha _c-\omega
   _c\right)\right)}{(m-1)^2}+\frac{\alpha _c \mu _{ci} (mx-1)+(x-1) \left(\omega _c-{\gamma}_i\right)}{m-1}&i\in b,
   \end{cases}\\
   N\Psi_i(x)&=-B_i+n_i(x)(\langle\alpha\rangle_{\mathbf{n}}(x)+\langle\omega\rangle_{\mathbf{n}}(x))\\
&=\begin{cases}
-\frac{\alpha_c\mu_{cc}(mx-1)+{\gamma}_a(1-x)}{M-1}-x\omega_c+x(\langle\alpha\rangle_{\mathbf{n}}(x)+\langle\omega\rangle_{\mathbf{n}}(x))&i=c\\
-\frac{\alpha_c\mu_{ci}(mx-1)+({\gamma}_i+\omega_c)(1-x)}{m-1} +\frac{(1-x)}{m-1}(\langle\alpha\rangle_{\mathbf{n}}(x)+\langle\omega\rangle_{\mathbf{n}}(x))&i\in b
\end{cases}\\
&=\begin{cases}
\frac{-m (x-1) x \bar{\alpha }-m (x-1) x \bar{\omega }+\alpha_c \mu _{cc}-m x \alpha _c \mu
   _{cc}+m x^2 \alpha _c + m x^2 \omega _c-m x \omega _c-x \alpha _c+(x-1) {\gamma }_c}{m-1}&i=c\\
   \frac{(x-1) \left(m (x-1) \bar{\alpha }+m (x-1) \bar{\omega }-(m x-1) \left(\alpha _c+\omega
   _c\right)\right)}{(m-1)^2}+\frac{\alpha _c \mu _{ci} (1-m x)+(x-1) \left(\omega _c+{\gamma}_i\right)}{M-1}&i\in b.
   \end{cases}
\end{align}
To find the resultant forces along the path we then take the inner product with $\mathbf{u}^{(c)}$.
Doing so gives
\begin{align}
   &f^{(c)}_\Phi(x)\nonumber\\
   &=\Phi_c(\mathbf{n}^{(c)}(x))u^{(c)}_c+\sum_{i\in b}\Phi_i(\mathbf{n}^{(c)}(x))u^{(c)}_b\nonumber\\
&=\sqrt{\frac{m-1}{m}}\Bigg[\frac{m (x-1) x \bar{\alpha }-m (x-1) x \bar{\omega }-\alpha _c \mu _{cc}+mx
   \alpha _c \mu _{cc}-m x^2 \alpha _c+m x^2 \omega _c-mx \omega _c+x \alpha
   _c-(x-1) {\gamma}_c}{m-1}\nonumber\\
   &\qquad-\frac{1}{m-1}\sum_{i\in b} \Bigg(-\frac{(x-1) \left(m (x-1) \bar{\alpha }+\bar{\omega } (m-mx)-(mx-1) \left(\alpha _c-\omega
   _c\right)\right)}{(m-1)^2}\nonumber\\
   &\qquad\qquad\qquad\qquad\qquad+\frac{\alpha _c \mu _{ci} (mx-1)+(x-1) \left(\omega _c-{\gamma}_i\right)}{m-1}\Bigg)\Bigg]\nonumber\\
   &=\sqrt{\frac{m-1}{m}}\Bigg[\frac{m (x-1) x \bar{\alpha }-m (x-1) x \bar{\omega }-\alpha _c \mu _{cc}+mx
   \alpha _c \mu _{cc}-m x^2 \alpha _c + m x^2 \omega _c - m x \omega _c+x \alpha
   _c-(x-1) {\gamma}_c}{m-1}\nonumber\\
   &\qquad- \Bigg(-\frac{(x-1) \left(m (x-1) \bar{\alpha }+\bar{\omega } (m-m x)-(m x-1) \left(\alpha _c-\omega
   _c\right)\right)}{(m-1)^2}\nonumber\\
   &\qquad\qquad\qquad\qquad\qquad+\frac{\alpha _c (1-\mu_{cc}) (m x-1)+(x-1) \left(\omega _c(m-1)-(m\bar{\alpha}-\gamma_c)\right)}{(m-1)^2}\Bigg)\Bigg]\nonumber\\
   &=\sqrt{\frac{m-1}{m}}\left[\frac{m \left((x-1) \left( mx \left(\bar{\alpha }-\bar{\omega }+\omega _c\right)+\bar{\omega }-\omega
   _c\right)-\alpha _c (m x-1) \left(t-\mu _{cc}\right)+\gamma _c-x \gamma _c\right)}{(m-1)^2}\right]\nonumber\\
     &=\sqrt{\frac{m}{(m-1)^3}}\left[(x-1) \left(m x \left(\bar{\alpha }-\bar{\omega }+\omega _c\right)+\bar{\omega }-\omega
   _c\right)-\alpha _c (m x-1) \left(x-\mu _{cc}\right)+\gamma _c(1-x)\right].
\end{align}
Similarly, we find
\begin{align}
   Nf^{(c)}_\Psi(x)&=\sqrt{\frac{m}{(m-1)^3}}\left[-(x-1) \left((m x-1) \left(\bar{\omega }-\omega _c\right)+m x \bar{\alpha }\right)+\alpha _c
   (m x-1) \left(x-\mu _{cc}\right)+(x-1) \gamma _c\right].
\end{align}
This then allows us to state the general form of the forces along, and in the direction of, the simplex bisection line as
\begin{empheq}[box=\fbox]{align}
   f^{(c)}(x)&=f^{(c)}_\Phi(x)+f^{(c)}_\Psi(x)+f^{(c)}_\Theta(x),\\
   f^{(c)}_\Phi(x)&=\sqrt{\frac{m}{(m-1)^3}}\left[(x-1) \left(m x \left(\bar{\alpha }-\bar{\omega }+\omega _c\right)+\bar{\omega }-\omega
   _c\right)-\alpha _c (m x-1) \left(x-\mu _{cc}\right)+\gamma _c(1-x)\right],\\
   Nf^{(c)}_\Psi(x)&=\sqrt{\frac{m}{(m-1)^3}}\left[-(x-1) \left((m x-1) \left(\bar{\omega }-\omega _c\right)+m x \bar{\alpha }\right)+\alpha _c
   (m x-1) \left(x-\mu _{cc}\right)+(x-1) \gamma _c\right],\\
   f^{(c)}_\Theta(x)&=\sqrt{\frac{m}{4N^2(m-1)^3}}\bigg[\alpha_c\left(m\mu_{cc}(mx+x-2)+(m+1)x(1-mx)\right)-\omega_c((m+1)m(x-1)x+m-1)\nonumber\\
   &\qquad\qquad\qquad\qquad\qquad+\bar{\alpha}(m-1)mx\mu_{cc}+(m+1)(x-1)(\bar{\alpha}mx-{\gamma}_c)+mx\bar{\omega}(mx+x-2)\bigg].
\end{empheq}
The total force along the bisecting line is then revealed as a quadratic function of the parameterisation variable $x$. Consequently, there are up to two fixed points, one stable and one unstable, which will control the qualitative behaviour of the system. We will explore this behaviour in the next sections.

\section{Fidelity catastrophe in the case of identically fit variants}
\label{sec:eq_variants}
In the first instance, a simple, but instructive, fitness landscape to consider for the purpose of illustrating the fidelity catastrophe is that of equally fit variants.
In this case we have $\alpha_i=\gamma_i=\bar{\alpha}=\alpha$ and $\omega_i=\bar{\omega}=\omega$, which results in a special case where the form of the forces reduce from a quadratic to a linear form. As in the preceding development, we assume only $\mu_{ij}=\mu_{ji}$ and $\mu_{ii}=\mu_{jj}=\hat{\mu}$ for all $i$ and $j$. Explicitly, this yields total force and components
\begin{equation}
f^{(c)}_\Phi(x) = \sqrt{\frac{m}{(m-1)^3}} \alpha(1-mx)(1-\hat{\mu}),
\end{equation}
\begin{equation}
f^{(c)}_\Psi(x) = \sqrt{\frac{m}{N^2(m-1)^3}}\alpha(mx-1)(1-\hat{\mu}),
\end{equation}
\begin{equation}
f^{(c)}_\Theta(x) = \sqrt{\frac{m}{4N^2(m-1)^3}}(mx-1)\Big[\alpha(m(2\hat{\mu}-1)-1)+(m-1)\omega \Big],
\label{eq:noiseF_eq_var}
\end{equation}
\begin{equation}
f^{(c)}(x)=\sqrt{\frac{m}{4N^2(m-1)^3}}\Big[(m x-1) \left[2 \alpha  (\hat{\mu} -1) N+\alpha  (2 \hat{\mu} -1) (m-1)+(m-1) \omega \right]\Big].
\end{equation}
Here there is only a single fixed point, trivially at $x=1/m$ (\emph{i.e.}, the centre of the simplex), and its location remains unchanged for all values of $\hat{\mu}$, $N$, $\alpha$ and $\omega$. The stability of this fixed point, however, can change.
\par
As there is no master variant in this fitness landscape, there is also no concept of an error catastrophe, however, the fidelity catastrophe can still be observed, and this is related to the stability of the sole fixed point. Since the forces are simple linear functions, the stability of the fixed point is simply related to the sign of the multiplying factor to the term in $(mx-1)$.
\par
\begin{figure}[!b]
   \centering
   \includegraphics[width=0.8\textwidth]{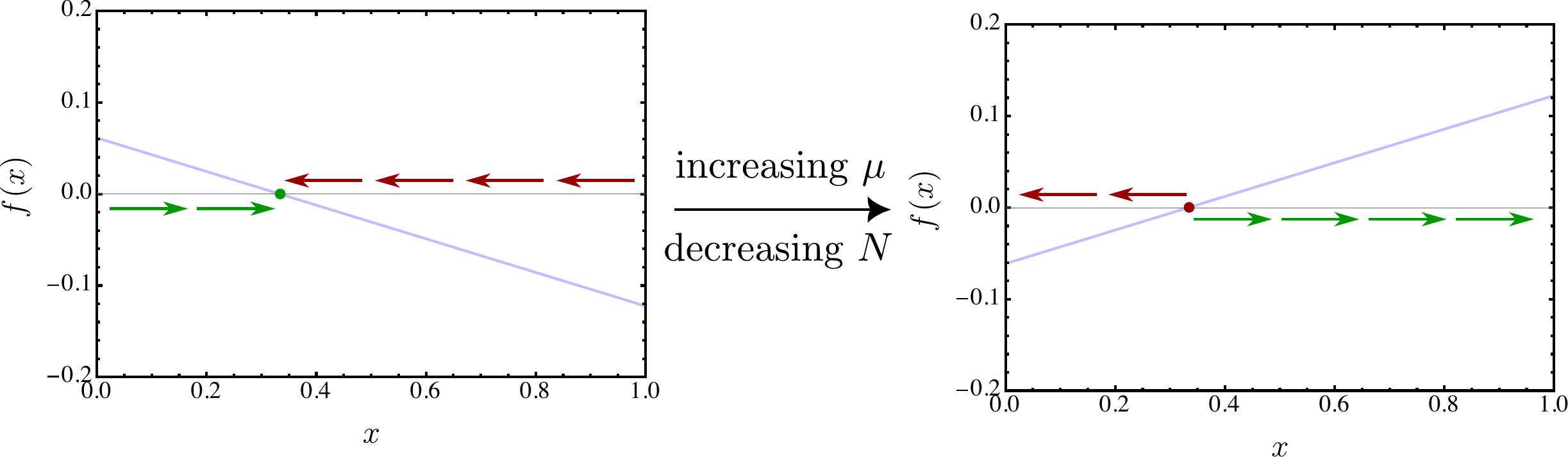} 
   \caption{Cartoon of the effective force experienced along a simplex bisection line in the case of equally fit variants. The force is a simple linear function with a fixed point at $x_*=1/m$ (shown here for $m=3$). As $\hat{\mu}$ is increased or $N$ is decreased the stability of this fixed point can change from stable (green point) to unstable (red point). When this happens the system is forced to the points $x=0$ and $x=1$ and the system's distributions changes from a uni-modal one centred at $x=1/m$ to a bi-modal one centred at $x=0$ and $x=1$. This is the fidelity catastrophe. When all such bisection lines are considered we understand that the total distribution is $m$-modal across the whole simplex.}
   \label{fig:eq_variants}
\end{figure}

Explicitly, when the pre-factor to the term in $(mx-1)$ is negative, the fixed point at $x=1/m$ is stable. As $\hat{\mu}$ is increased, or $N$ is decreased, the pre-factor can change sign, becoming positive, thus leaving the fixed point unstable. This is illustrated in Fig.~\ref{fig:eq_variants}.

When the fixed point is unstable, the distribution along the bisecting line becomes bimodal, with the edges of the simplex, namely $x=0$ and $x=1$ along all bisecting lines, become effective attractors. This is precisely what we observe in Fig.~\ref{fig:exact_2d} when we see the distribution become bimodal. In this case, when the pre-factor is precisely zero the force vanishes everywhere along the bisecting line. The expected behaviour of the system when not experiencing the fidelity catastrophe, such that fixed point is stable, is that of a well mixed system with approximately equal distribution between the variants. When the system experiences the fidelity catastrophe, such that the fixed point is unstable, we instead expect stochastic switching between temporary dominance of different single variants, with none being preferred, on average.
\par
We can easily characterise the onset of the fidelity catastrophe in this case by simply solving for when the total force vanishes (\emph{i.e.}, when the fixed point changes stability). Solving in terms of a threshold $N$ we have
\begin{align}
   N_{\rm crit}^{\rm fidel}&=\frac{(m-1)(\alpha(2\hat{\mu}-1)+\omega)}{2\alpha(1-\hat{\mu})}.
   \label{eq:eq_var_crit_N}
\end{align}
In general terms, changing the value of $N$ changes the relative weighting of the noise-induced force as compared to the deterministic contribution. The deterministic contribution acts to push the system into the centre of the simplex causing the fixed point at $x=1/m$ to act more like a stable fixed point, whilst, for sufficiently high $\hat{\mu}$, the noise-induced component acts to push the system to the edges of the simplex.
\par
Alternatively, solving in terms of a threshold fidelity, $\hat{\mu}$, yields
\begin{align}
   1-{\hat{\mu}}^{\rm fidel}_{\rm crit}&=\frac{(m-1)(\alpha+\omega)}{2\alpha(m+N-1)},
   \label{eq:eq_var_crit_mu}
\end{align}
revealing that, for suitably low $N$, a minimal fidelity is required to observe the switching phenomenon.
\par
Finally, we note that the onset of the switching behaviour arises due to the varied behaviour of the noise-induced force, which changes as a function of $\hat{\mu}$. Specifically, the noise-induced component itself (Eq.~\eqref{eq:noiseF_eq_var}) can act to either force the system into or away from the fixed point at $x=1/m$ depending on the value of $\hat{\mu}$ --- at low $\hat{\mu}$ the noise-induced force stabilises the system pushing the state towards the fixed point, however at high $\hat{\mu}$ the noise-induced force serves to push the system away from the fixed point.
\par
As such, we can consider the point where the noise-induced force vanishes as a function of $\hat{\mu}$, characterising when the noise-induced force changes its character from one that pushes inwards towards the centre of the simple, to one where it pushes outwards towards its edges. This is given by
\begin{align}
   1-{\hat{\mu}}_{\rm crit}^{\rm noise}&=\frac{m-1}{m}\cdot\frac{\alpha+\omega}{2\alpha},
   \label{eq:mu-inv}
\end{align}
which is conveniently a threshold independent of the other main control parameter, $N$. We note that in the case of approximately stable populations ($\alpha\sim \omega$), this threshold fidelity is a simple function of the size of the state space $\hat{\mu}_{\rm crit}^{\rm noise}\sim 1/m$ such that for almost any practical system (\emph{i.e.}, where $m\gg 1$) the threshold for the noise-induced force to direct the system away from the centre of the simplex becomes very low. We also note that we can use this threshold to empirically test our predictions of the nature of the noise-induced force which we perform in Sec.~\ref{sec:test_noise}.

We can then consider a more realistic genetic model, as per the main text, where genomes are made of $L$ bases in the set ${\cal A}$ ({\it e.g.}, ${\cal A}=\{\textrm{A,T,G,C}\}$), leading to a total of $m=|{\cal A}|^L$ variants connected by the mutation rate matrix
\begin{equation}
\label{eq:mut-ham}
\mu_{ij}=\left(\frac{\varepsilon}{|{\cal A}|-1}\right)^{H_{ij}}(1 - \varepsilon)^{L - H_{ij}},
\end{equation}
where $\varepsilon$ is the average error rate per base during replication and $H_{ij}$ is the Hamming distance between $i$ and $j$ --- \emph{i.e.}, the amount of loci where the genomes of $i$ and $j$ have different bases.
In this case we can identify $\hat{\mu}=(1-\varepsilon)^L$ and $m = |{\cal A}|^L$. This allows us to express the per base error rate that causes the noise-induced force to invert, \emph{viz.}
\begin{align}
\varepsilon^\textrm{noise}_\textrm{crit} &= 1 - \left( \frac{\alpha(|{\cal A}|^L+1)-\omega(|{\cal A}|^L-1)}{2\alpha|{\cal A}|^L} \right)^\frac{1}{L}\nonumber\\
&\simeq \frac{1}{L}\ln\left(\frac{2\alpha}{\alpha-\omega}\right) + {\cal O}(L^{-2})+{\cal O}(m^{-1}).
\end{align}

\section{Single dominant variant quasispecies case}

Next, we consider the case of a single dominant variant as often used in models of quasispecies. This is conveniently the simplest system parameterisation that retains the full quadratic character of the forces along the bisecting simplex line. We characterise such a situation by considering $m-1$ variants with replication and deterioration rate, $\alpha$ and $\omega$, respectively. Then we consider an additional single master variant, having index $z$, which has the same deterioration rate, $\omega$, but an elevated replication rate, $\alpha+\Delta \alpha$. The mutation rate $\mu_{ij}$ is arbitrary except for symmetry, $\mu_{ij}=\mu_{ji}$, and equal fidelity between the variants, \emph{i.e.}, $\mu_{ii}=\mu_{jj}=\hat{\mu}$ for all $i$ and $j$.
\par
In this case we have the summation variables
\begin{align}
   \bar{\omega}=\omega,\quad\bar{\alpha}=\alpha+\frac{\Delta\alpha}{m},\quad\gamma_i=\alpha+\Delta\alpha\mu_{zi},
\end{align}
in turn allowing us to calculate the forces along the bisection line. We note that we must be careful to distinguish the form of the forces depending on which bisection line of the simplex we choose.
\par
 First we consider the bisection line related to the vertex of the dominant species (such that $\alpha_c=\alpha_z=\alpha+\Delta\alpha$). This gives us forces
\begin{align}
   f^{(z)}_{\Phi}(x)&=\sqrt{\frac{m}{(m-1)^3}}\left[\alpha (1-m x) (1-\hat{\mu}) -\Delta \alpha  (m-1) x (x-\hat{\mu} )\right]\label{eq:f_comp_quasi1}\\
   f^{(z)}_{\Psi}(x)&=-N^{-1}f^{(z)}_{\Phi}(x)\label{eq:f_comp_quasi2}\\
   f^{(z)}_{\Theta}(x)&=\sqrt{\frac{m}{4N^2(m-1)^3}}\Big[ (m x-1) \left[\alpha \left(m \left(2 \hat{\mu}-1\right)-1\right)+(m-1) \omega\right]\nonumber\\
   &\qquad\qquad\qquad\qquad+\Delta\alpha (m-1) \left[((m+2) x -1)\hat{\mu}-(m+1) x^2\right]\Big].\label{eq:f_comp_quasi3}
   \end{align}
   On the other hand, forces along all other bisection lines (where $\alpha_c=\alpha$) are given by
   \begin{align}
   f^{(c\neq z)}_{\Phi}(x)&=\sqrt{\frac{m}{(m-1)^3}}\left[\alpha  (1-mx) (1-\hat{\mu})-\Delta \alpha  (1-x) (x-\mu_{zc})\right]\\
   f^{(c\neq z)}_{\Psi}(x)&=-N^{-1}f^{(c\neq z)}_{\Phi}(x)\\
   f^{(c\neq z)}_{\Theta}(x)&=\sqrt{\frac{m}{4N^2(m-1)^3}}\Big[(m x-1) \left[\alpha \left(m \left(2 \hat{\mu}-1\right)-1\right)+(m-1) \omega\right]\nonumber\\
   &\qquad\qquad\qquad\qquad+\Delta\alpha\left[(1 + m) (x-1) x + (m-1) x \hat{\mu} - (1 + m) (x-1) \mu_{zc}\right]\Big].
   \end{align}
Combining the deterministic and noise-induced forces along the bisecting line for the master and other variants gives parametrised total forces
\begin{align}
   \label{eq:forcez}
f^{(z)}(x)&=\sqrt{\frac{m}{4N^2(m-1)^3}}\left[(m x-1) \left[2 \alpha  (\hat{\mu} -1) N+\alpha  (2 \hat{\mu} -1) (m-1)+(m-1) \omega \right]\right.\nonumber\\
&\left.\qquad\qquad\qquad\qquad-\Delta\alpha  (m-1) \left(\hat{\mu} +x^2 (m+2 N-1)-\hat{\mu} 
   x (m+2 N)\right)\right],\\
   \label{eq:forcenotz}
 f^{(c\neq z)}(x)&= \sqrt{\frac{m}{4N^2(m-1)^3}}\left[ (m x-1) \left[2 \alpha  (\hat{\mu} -1) N+\alpha  (2 \hat{\mu} -1) (m-1)+(m-1) \omega \right]\right.\nonumber\\
 &\left.\qquad\qquad\qquad\qquad+\Delta \alpha  (m-1) \left(x \left(\hat{\mu} +x-\mu
   _{\text{zc}}-1\right)+\mu _{\text{zc}}\right)+2 \Delta \alpha  N (x-1) \left(x-\mu _{\text{zc}}\right)\right].
\end{align}
We note, setting $\Delta\alpha=0$, such that all species become equivalent, results in both expressions being equal (as they must), but also reduces the form of the forces to the linear expressions in $x$ found in Sec.~\ref{sec:eq_variants}, with a single fixed point at $x=1/m$.

Much of the behaviour of the quasispecies system can be understood through an appreciation of the forces that appear in Eqs.~\eqref{eq:forcez} and \eqref{eq:forcenotz}. First we consider the force along the bisecting line of the master variant which, on its own, is enough to characterise the nature of both the error and fidelity catastrophes. 
\par
The force along the master bisecting line is a simple quadratic force with a negative term in $x^2$ and as such we can understand the qualitative behaviour of the system in terms of the resultant, up to two, fixed points in the simplex where the force vanishes. Solving for vanishing force reveals fixed points
\begin{align}
    x_*^{(z)}&=\frac{1}{2\Delta\alpha(m-1)(2N+m-1)}(\nu\pm\sqrt{\zeta}),\\
    \nu&=\Delta\alpha \hat{\mu} (m-1) (m+2 N)+\alpha  m (2 (\hat{\mu} -1) N+(2 \hat{\mu} -1) (m-1))+(m-1) m \omega,\\
    \zeta&=\left(-2 \Delta\alpha \hat{\mu} N+m^2 (\alpha  (2 \hat{\mu} -1)+\Delta\alpha \hat{\mu} +\omega )+m (2 \alpha (\hat{\mu} -1) N-2 \alpha 
   \hat{\mu} +\alpha +\Delta\alpha \hat{\mu} (2 N-1)-\omega )\right)^2\nonumber\\
   &\qquad-4 \Delta\alpha  (m-1) (m+2 N-1) (2 \alpha  (\hat{\mu}-1)
   N+(m-1) (\Delta\alpha \hat{\mu} +\omega )+\alpha  (2 \hat{\mu} -1) (m-1)).
\end{align}
\par
Similarly, we may consider the force along the bisecting line of a variant which is not the master. This too is a quadratic, but has a positive $x^2$ term. Solving for its vanishing fixed points yields
\begin{align}
   x_*^{(c\neq z)}&=\frac{1}{2\Delta\alpha(2N+m-1)}(\nu\pm\sqrt{\zeta}),\\
   \nu&=m^2 (-2 \alpha \hat{\mu} +\alpha -\omega )+m \left(\alpha  (-2 \hat{\mu}  N+2 \hat{\mu} +2 N-1)+\Delta \alpha  (-\hat{\mu} )+\Delta \alpha +\omega
   +\Delta \alpha  \mu_{\text{zc}}\right)\nonumber\\
   &\qquad+\Delta \alpha  \left(\hat{\mu} +2 N+(2 N-1) \mu _{\text{zc}}-1\right)\\
   \zeta&=\Big[\Delta \alpha  (\hat{\mu} +2 N-1)+m^2 (-2 \alpha \hat{\mu} +\alpha -\omega )+m (\alpha  (-2 \hat{\mu}  N+2 \hat{\mu} +2 N-1)+\Delta \alpha 
   (-\hat{\mu})+\Delta \alpha +\omega )\nonumber\\
   &\qquad\qquad+\Delta \alpha  (m+2 N-1) \mu_{\text{zc}}\Big]^2\nonumber\\
   &\quad-4 \Delta \alpha  (m+2 N-1)
   \left(\alpha  (-2 \hat{\mu}  N+2 \hat{\mu} -2 \hat{\mu}  m+m+2 N-1)-m \omega +\Delta \alpha  (m+2 N-1) \mu _{\text{zc}}+\omega \right).
\end{align}
First, we turn our attention to the behaviour of the force along bisection line of the master variant.
\par
\begin{figure}[!t]
   \centering
   \includegraphics[width=0.5\textwidth]{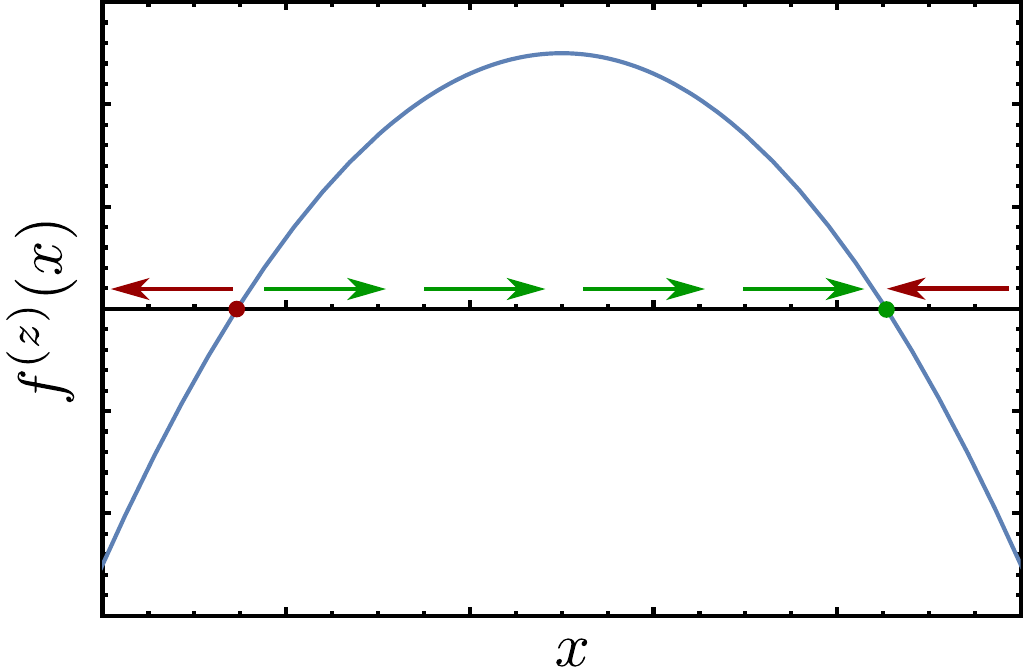} 
   \caption{Cartoon of the effective force experienced along the simplex bisection line of the master variant. The form is a negative quadratic, thus there are two potential fixed points. A stable one (green point) which acts as an attractor, and an unstable one (red point) which repels the system. In general, the stable and unstable fixed points are bounded by $x>0$ and $x<1$ (the lower and upper physical limits of the system), respectively, but each point need not necessarily lie inside the simplex --- \emph{i.e.}, the unstable fixed point can lie below $x=0$ and the stable fixed point can lie above $x=1$. If the unstable fixed point lies within the simplex (\emph{i.e.}, for $x>0$), then the master variant extinction state ($x=0$) effectively acts as an attractor, destabilising unimodal quasi-deterministic solutions, whereas the system is repelled from the master extinction if the fixed point lies below $x=0$, stabilising them. In contrast, the location  of the stable fixed point determines the behaviour of the system which comports with the deterministic Eigen model.}
   \label{fig:quad}
\end{figure}

In this case, because the quadratic pre-factor is negative, the upper of these fixed points is stable and associated with the concentration of probability around the dominant variant in the sense of the deterministic Eigen model. The lower fixed point is unstable and is broadly associated with noise-induced forces.
The overall behaviour induced by the quadratic force, and the relevance of the fixed points, is shown in Fig.~\ref{fig:quad}.

The existence of the unstable fixed point marks a significant deviation from the behaviour of the deterministic Eigen model. When it exists within the simplex, a large enough fluctuation (\emph{i.e.}, to a low enough value of $x$) will cause the parameterised state of the system to run away to $x=0$ characterising relative depletion of the master variant, such that the system is temporarily dominated by a variant which is not the master. This behaviour occurs only in cases when the fidelity, $\hat{\mu}$, is above a certain threshold, as broadly observed in the equal two variant case in Fig.~\ref{fig:exact_2d} when the distribution becomes bimodal. Consequently, as with the equal variant case in the previous section, we call this behaviour the fidelity catastrophe, and we consider the threshold of this behaviour to occur at the value of $\hat{\mu}$ where the unstable fixed point enters the simplex, $\hat{\mu}_{\rm crit}^{\rm fidel}$. 
\par
In contrast, the upper, stable, fixed point corresponds to the expected behaviour of the deterministic model, serving as an attractor of the system associated with dominance of the master variant. Here, decreases in fidelity (increased error) disrupt the dominance of the master species, causing the stable fixed point to reside at progressively lower values of $x$. When this reduction reaches a threshold level, $\hat{\mu}_{\rm crit}^{\rm error}$, we say that the system has suffered from the error catastrophe.
\par
We characterise these thresholds in the subsequent sub-sections.

\subsection{Fidelity catastrophe}
\par 
We first consider the fidelity catastrophe. This occurs when the fidelity, $\hat{\mu}$, is too high, or equivalently, when the ratio $m/N$ is too low and causes edges of the simplex to become attractors.
The signature of this behaviour experienced in the force along the master species bisecting line is the entry into the simplex of the unstable fixed point from outside the simplex, at $x=0$.
This then causes the system in configurations near $x=0$ to experience a force that drives them towards extinction. We can characterise this very simply by solving for vanishing force along the bisecting line at $x=0$, \emph{i.e.}, $f^{(c)}(0)=0$.
Solving for this condition in terms of $\hat{\mu}$ reveals a threshold fidelity
\begin{align}
    \hat{\mu}^{\rm fidel}_{\rm crit}&=\frac{\alpha  (m+2 N-1)-m \omega +\omega }{\Delta\alpha  (m-1)+2 \alpha  (m+N-1)}
\end{align}
or alternatively a threshold system size
\begin{align}
    N^{\rm fidel}_{\rm crit}&=\frac{(m-1) (\alpha  (2 \hat{\mu} -1)+\Delta\alpha \hat{\mu} +\omega )}{2 \alpha  (1-\hat{\mu})}.
\end{align}
Note this definition subsumes the definitions introduced in the equal variants case since setting $\Delta \alpha=0$ recovers Eqs.~\eqref{eq:eq_var_crit_N} and \eqref{eq:eq_var_crit_mu}.
\subsection{Error catastrophe}
Since this quasispecies fitness landscape possesses a well defined master variant it is instructive to consider the manifestation of the classic error catastrophe or error threshold in this system.
\par
This fixed point does not have as simply defined a threshold as with the fidelity catastrophe. Instead, we note that for appropriate system parameters, as $\hat{\mu}$ is reduced (error is increased), the system experiences a rapid change between a regime of approximately linear response in the position of $x_*^{(z)}$ to a regime where the position of the fixed point is relatively independent of $\hat{\mu}$ near $x=1/m$. We choose to characterise this crossover behaviour in terms of a threshold through the point of maximal curvature in the position of the fixed point with respect to $\hat{\mu}$, \emph{i.e.}, $d^3x_*^{(z)}(\hat{\mu})/d\hat{\mu}^3=0$. Since the fixed point is the solution of a quadratic this has an analytical solution, and is given by
\begin{align}
   \hat{\mu}_{\rm crit}^{\rm error}&=\frac{A}{B},\\
   A&=\Delta \alpha  (m-1) \left(\alpha  (m+2 N-1) \left(m^2+2 (m+2) N+4 m-4\right)-(m-1) m \omega  (m+2 N)\right)\nonumber\\
   &\qquad+2 \alpha 
   m^2 (m+N-1) (\alpha  (m+2 N-1)-m \omega +\omega )+2 (\Delta \alpha) ^2 (m-1)^2 (m+2 N-1),\\
   B&=(2 \alpha  m (m+N-1)+\Delta
   \alpha  (m-1) (m+2 N))^2.
\end{align}
Such an expression amounts to a powerful generalisation of the original heuristic formula for the error threshold, containing both population and phase space/genome size dependence through $N$ and $m$, respectively. As such the location of the threshold can deviate significantly from that originally postulated by Eigen if $N$ is not large enough compared to $m$ to effect a deterministic limit or generally if $m$ is independently not large. However, we can easily obtain the original result by progressively considering the deterministic, and large $m$, limits. First we expand in orders of $N^{-1}$ to obtain
\begin{align}
    \hat{\mu}_{\rm crit}^{\rm error}&=\frac{\alpha  (m (\Delta \alpha +m (\alpha +\Delta \alpha ))-2 \Delta \alpha )}{(\Delta \alpha -m (\alpha +\Delta \alpha
   ))^2}+\mathcal{O}(N^{-1}),
\end{align}
thus revealing the location of the error threshold in the deterministic, but small state space, limit. By then expanding in powers of $m^{-1}$ we find
\begin{align}
\hat{\mu}_{\rm crit}^{\rm error}&=\frac{\alpha}{\alpha+\Delta\alpha}+\frac{3\alpha\Delta\alpha}{(\alpha+\Delta\alpha)^2}\frac{1}{m}+\mathcal{O}(m^{-2})+\mathcal{O}(N^{-1}),
\label{eq:err_thresh_classic}
\end{align}
agreeing with the classical result \cite{eigen1971selforganization,eigen1977principle} as $m\to\infty$.
\par
Note, however that this agreement holds in the limit $N\gg m\gg 1$ only. If we exchange the order of the limits such that we consider $m\gg N$, we instead obtain
\begin{align}
\hat{\mu}_{\rm crit}^{\rm error}&=\frac{\alpha-\omega}{2\alpha+\Delta\alpha}+\frac{2 \left(\alpha ^2 N+N \omega  (\alpha +\Delta \alpha )+2 \alpha  \Delta \alpha
   +(\Delta \alpha) ^2\right)}{(2 \alpha +\Delta \alpha )^2}\frac{1}{m}+\mathcal{O}(m^{-2})
\end{align}

\subsection{Relationship between the error and fidelity catastrophes}

Moreover, in the limit $m\gg N$ these two error thresholds converge on each other. Specifically, we find
\begin{align}
\frac{\hat{\mu}_{\rm crit}^{\rm fidel}}{\hat{\mu}_{\rm crit}^{\rm error}}&=1+\frac{2\Delta\alpha}{2\alpha+\Delta\alpha}\frac{N}{m}+\mathcal{O}(m^{-1}).
\label{eq:error_ratio}
\end{align}
Consequently, in this highly relevant limit, $m\gg N$, there is a fundamental trade-off between these two (potentially) deleterious behaviours. As such, unless the number of replicators can be taken large enough that it exceeds the number of variants, assuming that the error catastrophe must be avoided in nature, we expect to see noise-induced effects causing the master species to be unstable over time.

\subsection{Illustration of key behaviours}

We now illustrate the main behaviour of the master variant fixed points and their relation to the error and fidelity catastrophe thresholds.
\par
\begin{figure}[!b]
   \centering
   \includegraphics[width=0.99\textwidth]{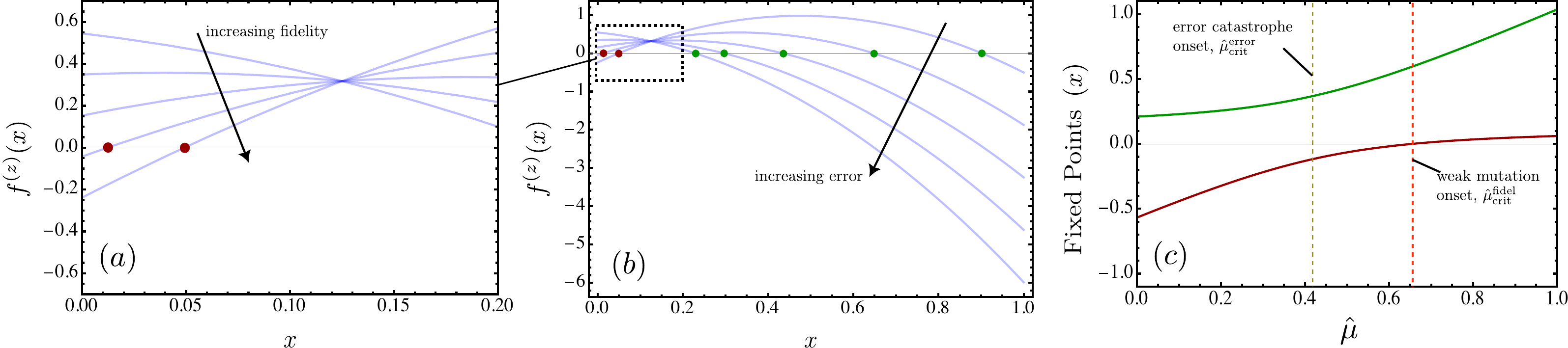} 
   \caption{Effective force experienced along the simplex bisection line of the master variant for a system with $m=3$, $N=10$, $\alpha=1$, $\Delta\alpha=4$, and $\omega=1/2$. Blue lines are the effective force along the simplex bisecting line and vary between fidelity parameter, $\hat{\mu}$, taking values from $0.1$ to $0.9$ in steps of $0.2$. As the fidelity is increased the lower fixed point (red dots) enters the simplex such that the force at $x=0$ becomes negative. The behaviour of this fixed point is shown by the red line in panel (c) with the onset of the fidelity catastrophe defined as the point where it enters the simplex at $x=0$. On the other hand, as the fidelity is reduced (error is increased), the upper, stable, fixed point (green dots) moves to lower values of $x$ in the simplex. The behaviour of this fixed point is shown by the green line in panel (c) with the onset of the error catastrophe defined by the point of maximum curvature in the fixed point position with $\hat{\mu}$. }
   \label{fig:fixed_points}
\end{figure}

In Fig.~\ref{fig:fixed_points} we illustrate the position of the fixed points with varying fidelity, $\hat{\mu}$, for an illustrative set of system parameters. In the left two sub panels are the effective forces along the bisecting line of the simplex for the master variant, with fixed points marked as red and green dots. The behaviour of the position of these fixed points is then plot in the right most sub panel. As $\hat{\mu}$ is increased (or equivalently error is reduced), we observe the lower fixed point enter the simplex (red dots), causing total depletion of the master variant to become a stable fixed point associated with the fidelity catastrophe. Conversely, as we reduce the fidelity (increase the error) the upper fixed point moves to lower and lower positions (values of $x$) within the simplex, causing relative depletion of the master variant, eventually associated with the error catastrophe.
\par
\begin{figure}[!t]
   \centering
   \includegraphics[width=0.99\textwidth]{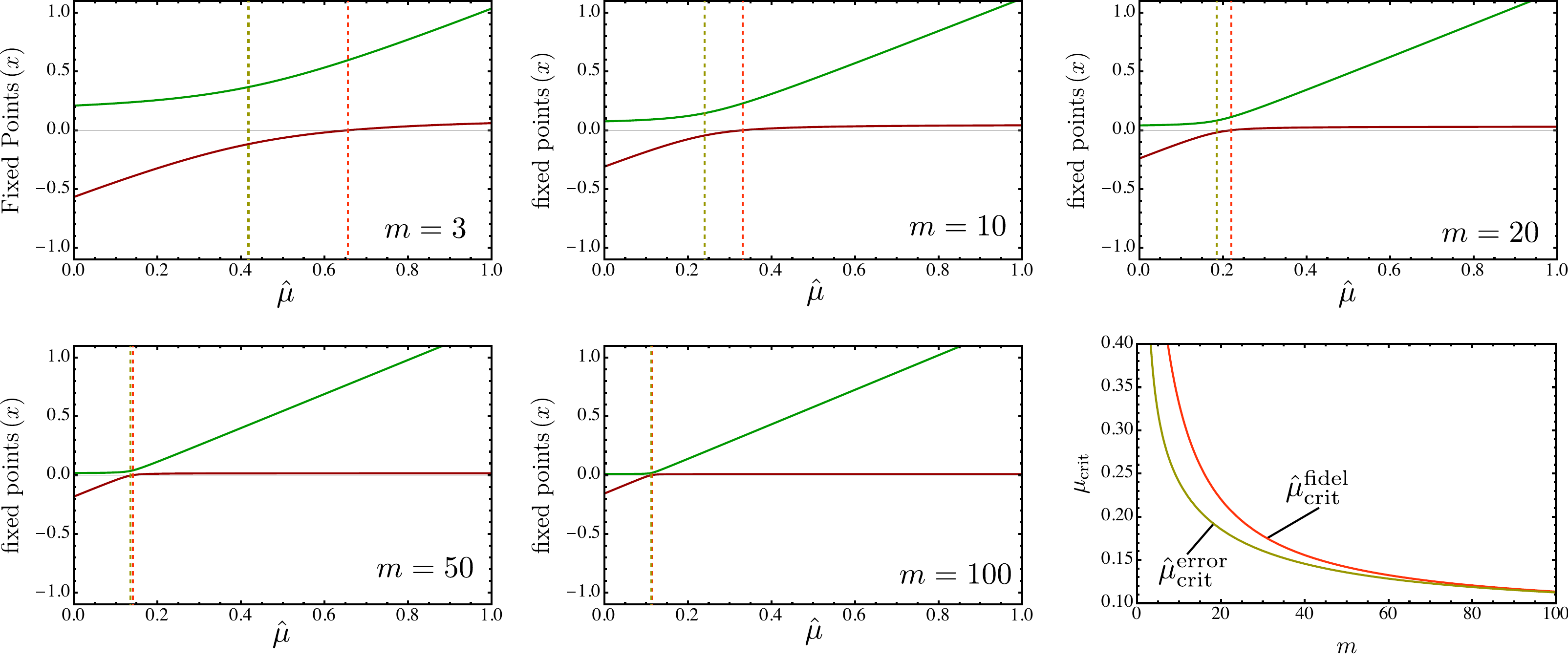} 
   \caption{Behaviour of the fixed points with $\hat{\mu}$, and related catastrophe thresholds, for different values of $m$. Keeping $N$ fixed for a system with parameters $N=10$, $\alpha=1$, $\Delta\alpha=4$, and $\omega=1/2$, increasing $m$ increases the ratio $m/N$ reducing the number of replicators per variant. From top left, through top right, to lower centre the location of the fixed points are shown for increasing $m$ from $m=3$ to $m=100$. Green lines indicate the position of the upper, stable, fixed point, whilst the red lines indicate the position of the lower, unstable, fixed point. The sub panel in the lower right shows the position of the onset of the error catastrophe (green) and fidelity catastrophe (red) as a function of $m$. As $m$ increases the position of the thresholds moves to lower values of $\hat{\mu}$. Moreover, the location of the thresholds tend closer to each other, reducing the region of error/fidelity space, $\hat{\mu}$, where one of the phnomena is not present. Further, as $m$ increases, the behaviour around the thresholds can be seen to get sharper, with the error catastrophe, in particular, becoming a more clearly resolved phenomenon. We note, that at large $m$ the centre of the simplex is at lower values of $x$ (\emph{i.e.}, at $x\sim 1/m$) such that it is becomes harder to see the fidelity catastrophe on the linear scales shown in the above plots.}
   \label{fig:vary_m}
\end{figure}

In Fig.~\ref{fig:vary_m} we illustrate the effect of varying the state space of the system (number of variants, $m$) on the positions of the fixed point within the simplex. Shown are positions of the fixed points varying with $\hat{\mu}$ for different values of $m$. Here, we see the resolution of the error catastrophe become much clearer, with the vertical dashed green lines increasingly separating distinct behaviour as $m$ increases. Moreover, we can also clearly see the prediction of Eq.~\eqref{eq:error_ratio} with the two thresholds tending to each other.
\par
\par
\begin{figure}[!h]
   \centering
   \includegraphics[width=0.99\textwidth]{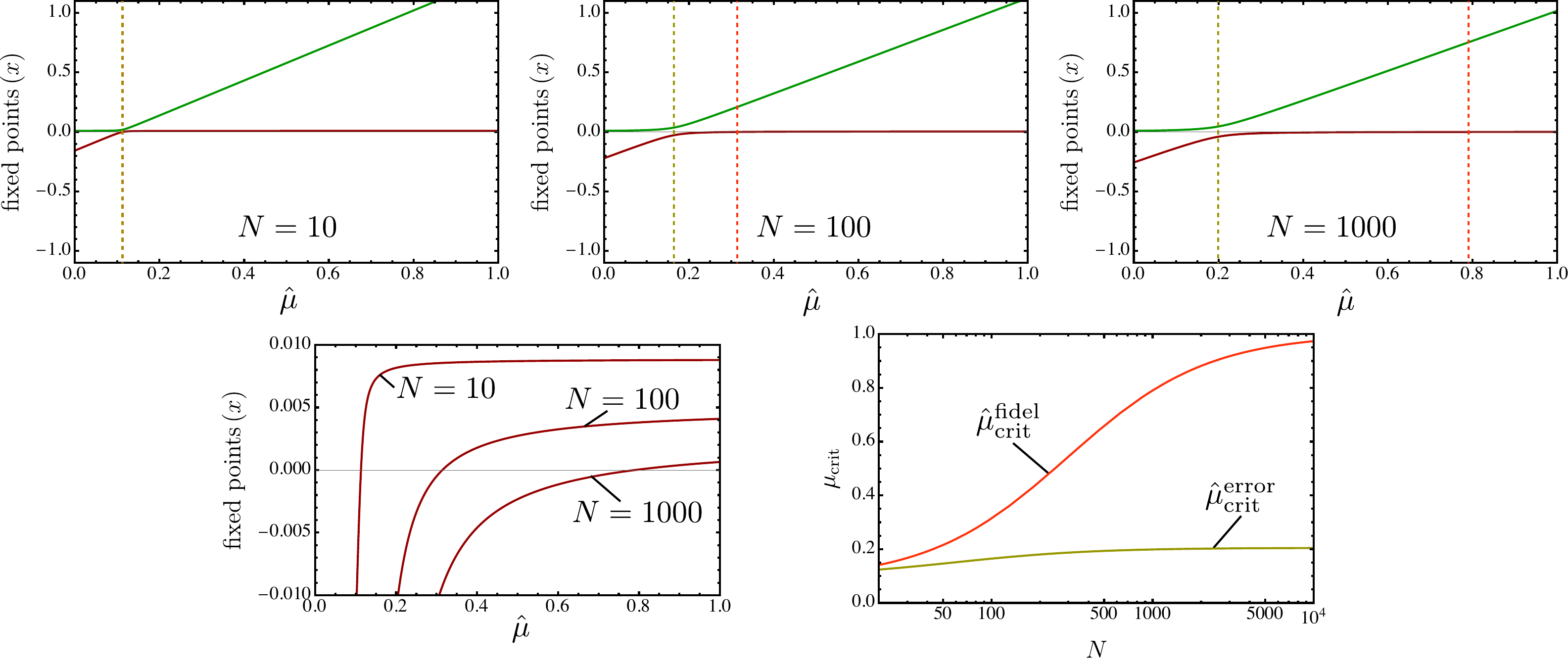} 
   \caption{Behaviour of the fixed points with $\hat{\mu}$, and related catastrophe thresholds, for different values of $N$. Keeping $m$ fixed for a system with parameters $m=100$, $\alpha=1$, $\Delta\alpha=4$, and $\omega=1/2$, increasing $N$ decreases the ratio $m/N$ increasing the number of replicators per variant. From top left to top right, the location of the fixed points are shown for increasing $N$ from $N=10$ to $N=1000$. Green lines indicate the position of the upper, stable, fixed point, whilst the red lines indicate the position of the lower, unstable, fixed point. The sub panel in the lower left shows a zoomed in view of the position of the lower fixed point for all values of $N$. The sub panel in the lower right shows the position of the onset of the error catastrophe (green) and fidelity catastrophe (red) as a function of $N$ on a logarithmic scale. As $N$ increases the position of the thresholds moves to higher values of $\hat{\mu}$. Moreover, the location of the thresholds tend away from each other, increasing the region of error/fidelity space, $\hat{\mu}$, where one of the phenomena (error catastrophe and weak mutation regime) is not present. In particular, as $N$ increases far beyond the value of $m$, such that there are many replicators per variant, the fidelity required to achieve the fidelity catastrophe tends to $1$. We also note that for $N\gg m$ and $m\gg 1$, most closely replicated by the top right panel ($N=1000$) the error threshold tends to the classical result $\alpha/(\alpha+\Delta\alpha)=0.2$.}
   \label{fig:vary_N}
\end{figure}

Finally, in Fig.~\ref{fig:vary_N} we illustrate the effect of varying the number of replicators in the system, $N$, on the positions of the fixed point within the simplex. Shown are positions of the fixed points varying with $\hat{\mu}$ for different values of $N$ at fixed $m$. Here, we see that the onset of the fidelity catastrophe is clearly a stochastic effect, with larger and larger fidelities (tending to $1$) required to observe the effect as $N$ is taken larger. We also note that in this figure (specifically in the top right sub panel) we consider  the model in the approach to the joint limit $N\gg m$ and $m\gg1$ required for it to replicate the deterministic limit of the large state space regime. As such the error threshold here can be seen to reside at $\sim0.2$, exactly at the value predicted by Eq.~\eqref{eq:err_thresh_classic} and which relates to the classic error threshold.
\par
In totality, the central observation that underlies the behaviour is the importance of the ratio $m/N$. When $m/N$ is small there are a large number of replicators per variant and thus stochastic effects are small. In this limit the fidelity catastrophe only occurs for extremely large values of $\hat{\mu}$ ($\sim 1$) and generally does not occur in the vicinity of the error catastrophe. In addition, if the state space is independently large ($m\gg 1$), then the error threshold coincides with the conventional limit associated with the Eigen model, $\alpha/(\alpha+\Delta\alpha)$. In contrast, when $m/N$ is large there are, on average, very few replicators per variant (\emph{i.e.}, significantly less than one) and thus stochastic effects are highly significant. In this limit the error and fidelity thresholds coincide such that it becomes practically impossible not to observe one of the error catastrophes for a given set of system parameters. It is then not irrelevant that for realistic models of genomes, the number of variants $m$ will be extremely large. It is instructive to then consider the  role and relevance of the assumptions that underpin the traditional, deterministic, Eigen model, which, by ignoring stochastic effects, essentially assumes $N\gg m$.

We have established thresholds for the fidelity and error catastrophes from the total force experienced by the master variant, however this only established values of $\hat{\mu}$ and $N$ where an extinction force arises for the master variant and where the position of the stable fixed point for the master variant qualitatively changes behaviour, respectively. It is instructive, therefore, to also examine the behaviour of the non-master variants.
The nature of the force on the non-master variants is inherently more complicated since it depends on the mutation fraction $\mu_{zc}$. These values, in concert with the other parameters, will control the overall qualitative behaviour of the system in a non-trivial way, however we can still find analogous thresholds in the behaviour of these non-master variants.
\par
First, we consider behaviour in a non-master variant that is analogous to the fidelity catastrophe. Here, at large enough $\hat{\mu}$ and low enough $N$, this variant, which is not expected to be dominant in the deterministic Eigen model, can develop an attractor at $x=1$ (\emph{i.e.}, population dominance) when the upper, unstable, fixed point enters the simplex at $x=1$. This is because the form of the force along this bisection line is a quadratic with positive curvature. Moreover, by defining a threshold for this behaviour in terms of when this fixed point enters the simplex (by solving for vanishing force at $x=1$) we find a result that is independent of $\mu_{zc}$. Solving in terms of a threshold $\hat{\mu}$ and $N$ here yields
\begin{align}
    \hat{\mu}^{\rm wm}_{{\rm crit},\neq z}&=\frac{\alpha(m+2N-1)+\omega-m\omega}{\Delta\alpha +2\alpha(m+N-1)},
\end{align}
and
\begin{align}
    N^{\rm wm}_{{\rm crit},\neq z}&=\frac{\Delta \alpha  \hat{\mu} +\alpha  (2 \hat{\mu} -1) (m-1)+(m-1) \omega }{2 \alpha  (1 - \hat{\mu} )}.
\end{align}
These expressions are very similar to, but not exactly equal to the earlier expressions for $\hat{\mu}_{\rm crit}^{\rm fidel}$ and $N_{\rm crit}^{\rm fidel}$, differing by a factor of $(m-1)$ into terms in $\Delta \alpha$ in the denominator and numerator, respectively. As a consequence, these thresholds occur at larger and smaller values of $\hat{\mu}$ and $N$, respectively.
\par
Similarly, behaviour in the lower, stable, fixed point exhibits a response which acts like an analogue to the error catastrophe. Specifically, as $\hat{\mu}$ is decreased (such that error is increased) the fixed point can also enter the simplex, such that extinction in the non-master variant can cease to locally be the most stable configuration of the system. We can solve for when this occurs yielding a threshold
\begin{align}
    \hat{\mu}_{{\rm crit},\neq z}^{{\rm error}}&=\frac{\alpha  (m+2 N-1)-m \omega +\omega+\Delta \alpha  (m+2 N-1) \mu _{\text{zc}} }{2 \alpha  (m+N-1)},
    \label{eq:mu_crit_fidel}
\end{align}
and
\begin{align}
    N_{{\rm crit},\neq z}^{{\rm error}}&=\frac{(m-1) \left(\alpha  (2 \hat{\mu} -1)+\omega -\Delta \alpha  \mu _{\text{zc}}\right)}{2 \left(\alpha (1-\hat{\mu})  +\Delta
   \alpha  \mu _{\text{zc}}\right)}.
   \label{eq:N_crit_fidel}
\end{align}
These thresholds, however, depend on the mutation fraction $\mu_{zc}$.
\par
\begin{figure}[!h]
   \centering
   \includegraphics[width=0.9\textwidth]{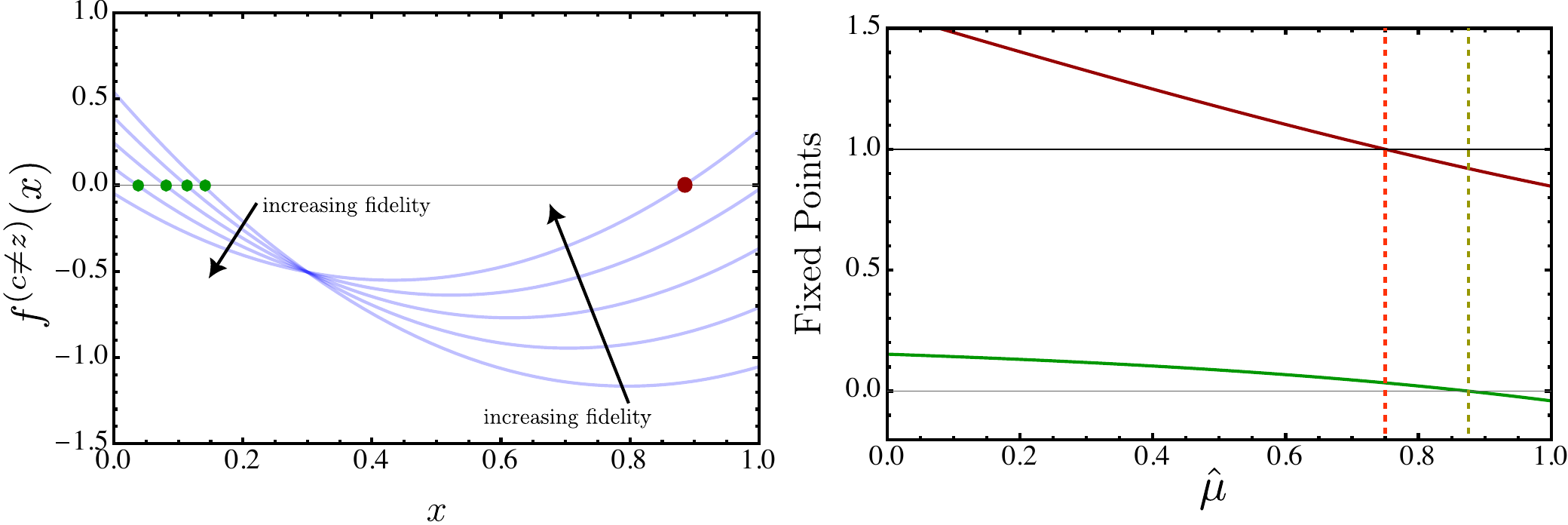} 
   \caption{Effective force experienced along the simplex bisection line of a non-master variant, $c$, for a system with $\mu_{zc}=0$, $m=3$, $N=10$, $\alpha=1$, $\Delta\alpha=4$, and $\omega=1/2$. Blue lines are the effective force along the simplex bisecting line and vary between fidelity parameter, $\hat{\mu}$, taking values from $0.1$ to $0.9$ in steps of $0.2$. As the fidelity is increased the upper fixed point (red dots) enters the simplex such that the force at $x=1$ becomes positive. The behaviour of this fixed point is shown by the red line in the right sub-panel with the threshold where it enters the simplex at $x=1$ shown by the vertical red line. As fidelity is increased further, the lower, stable, fixed point (green dots) moves out of the simplex, such that extinction becomes a stable fixed point. The behaviour of this fixed point is shown by the green line in the right sub-panel with the threshold shown by the vertical green line. }
   \label{fig:non_master}
\end{figure}

The behaviour of the force for a non-master variant, alongside the behaviour of the fixed points and their relation to the above thresholds is shown in Fig.~\ref{fig:non_master}.

\section{Empirical evidence for the noise-induced force}
\label{sec:test_noise}
\begin{figure}[!t]
   \centering
   \includegraphics[width=0.7\textwidth]{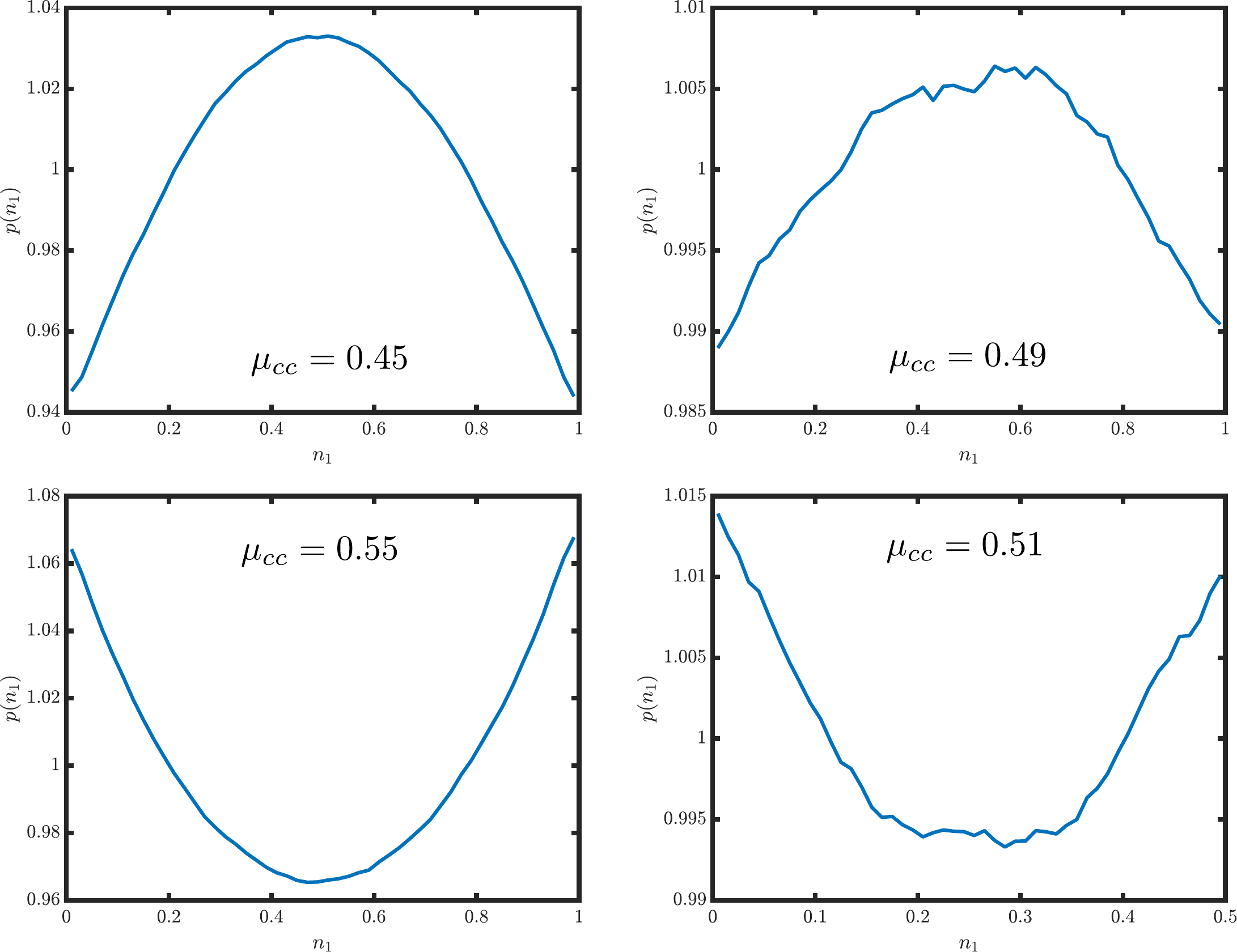}
   \caption{Empirical probability density functions for the system comprised only of stochastic terms for $m=2$, $\alpha_i=\omega_i=1$, for all $i$, and $\mu_{11}=\mu_{22}=\hat{\mu}$. The behaviour inverts at $\hat{\mu}=1/2$, as predicted by Eq.~\eqref{eq:mu-inv}.}
   \label{fig:empirical_2d}
\end{figure}

To confirm the qualitative behaviour of the noise-induced forces that underpin the overall behaviour, we can simulate the noise terms for the system in a trivial case and show that it possesses the same behaviour predicted by Eq.~\eqref{eq:noise-force-line}. Specifically, we simulate the SDEs in the case $m=2$ and $\alpha_i=\omega_i=1$, for all $i$, but remove the deterministic component to fully isolate the noise-induced behaviour. In this case the SDEs reduce to
\begin{align}
   dn_1&=\sum_{j=1}^2 q_{1j}dW_j\nonumber\\
   &=(1-n_1)\sqrt{B_1}dW_1-n_1\sqrt{B_2}dW_2,\\
   dn_2&=\sum_{j=1}^2 q_{2j}dW_j\nonumber\\
   &=(1-n_2)\sqrt{B_2}dW_2-n_2\sqrt{B_1}dW_1,
\end{align}
where
\begin{align}
   B_1=\sum_{j=1}^2n_j\alpha\mu_{j1}+n_1\omega=n_1\alpha\mu_{11}+n_2\alpha\mu_{21}+n_1\omega,\\
   B_2=\sum_{j=1}^2n_j\alpha\mu_{j2}+n_2\omega=n_2\alpha\mu_{22}+n_1\alpha\mu_{12}+n_2\omega.
\end{align}
As per the assumptions of the model, and dimension of the current example, we then have $\mu_{11}=\mu_{22}=\hat{\mu}$ and $\mu_{12}=\mu_{21}=1-\hat{\mu}$. For these values Eq.~\eqref{eq:noise-force-line} simplifies to
\begin{align}
   Nf_{\Theta(x)}^{(c)}(x)&=2\sqrt{2}(1-2x)(1-2\hat{\mu}).
\end{align}
Crucially, this predicts that the force has a zero at $x=1/2$, \emph{i.e.}, the fully mixed state $n_1=n_2=1/2$, and, moreover, the direction of this force inverts from one that pushes towards this centre point, to one that push outwards from this point, at a value of $\hat{\mu}=1/2$, also in agreement with Eq.~\eqref{eq:mu-inv}. Simulating the above SDEs under the additional stipulation of reflecting boundaries exactly corroborates this behaviour as shown in the probability density functions of system state shown in Fig.~\ref{fig:empirical_2d}, indicating that it has been formulated correctly.
\par
\begin{figure}[!b]
   \centering 
   \includegraphics[width=0.9\textwidth]{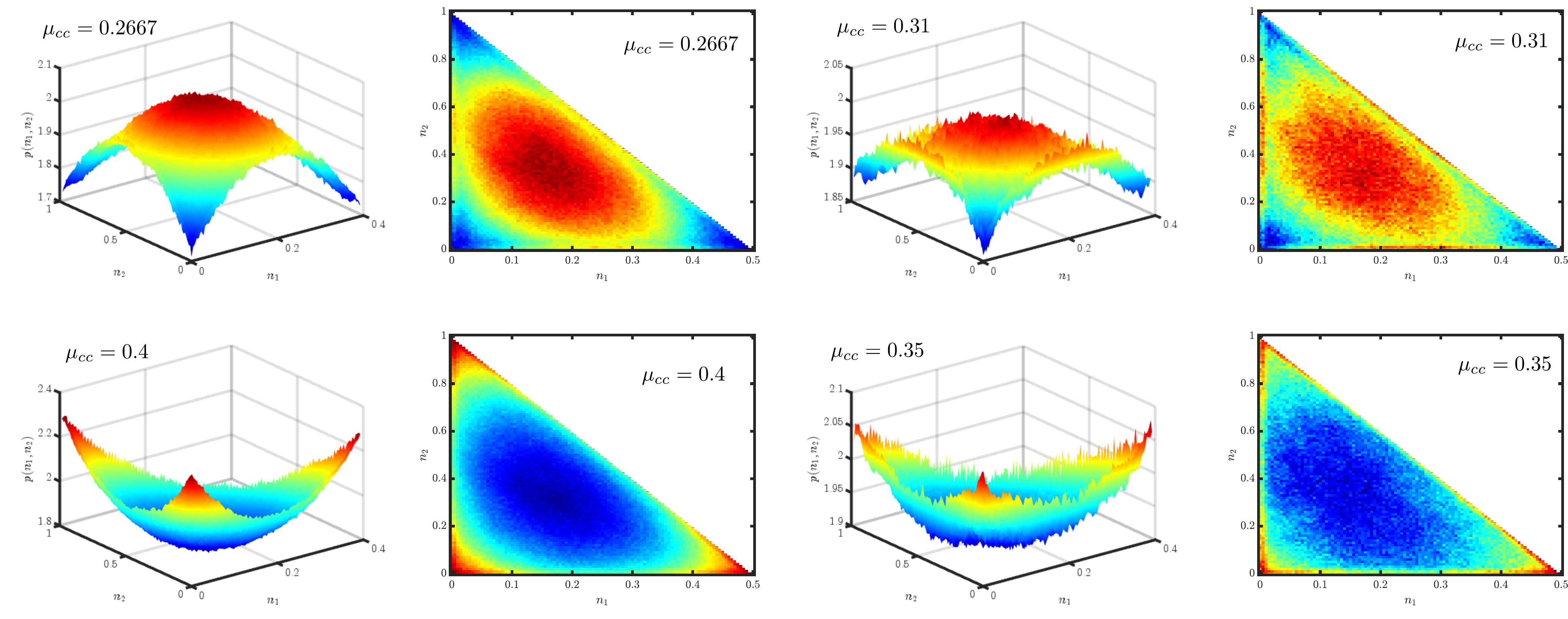}
   \caption{Empirical probability density functions for the system comprised only of stochastic terms for $m=3$, $\alpha_i=\omega_i=1$, for all $i$, and $\mu_{11}=\mu_{22}=\mu_{33}=\hat{\mu}$ and $\mu_{ij}=(1-\hat{\mu})/2$ for all $i\neq j$. The behaviour inverts at $\hat{\mu}=1/3$, as predicted by Eq.~\eqref{eq:mu-inv}.}
   \label{fig:empirical_3d}
\end{figure}

Finally, for completeness we offer empirical evidence for the form of the noise-induced force in the slightly less trivial case of $m=3$. Here Eq.~\eqref{eq:mu-inv} predicts that such an inversion (for the same parameters) occurs at $\hat{\mu}=1/3$. Again, we see exact corroboration of this behaviour, as seen in the probability density functions, for the system comprising only the stochastic terms, shown in Fig.~\ref{fig:empirical_3d}.

\section{Simulation}
Here we discuss numerical simulation of the underlying system
\begin{subequations}%
\begin{align}%
N_i &\xrightarrow{\sum_jN_j\alpha_j\mu_{ji}} N_i + 1,\\
N_i &\xrightarrow{N_i\omega_i} N_i - 1,
\end{align}%
\end{subequations}%
which underpin the master equation in Eq.~(\ref{eq:master}), that has formed the basis of our analysis.
\par
We simulate the system starting with $N(t=0)=N=\sum_{i=1}^m N_i(t=0)$ individuals at time $t=0$ through the use of the Gillespie algorithm, wherein events occur at intervals in time according to an exponential distribution with rate $\Lambda=\sum_{i=1}^m N_i(t)\alpha_i+N_i(t)\omega_i$, where $\alpha_i$ and $\omega_i$ are the replication and deterioration rates of variant $i$, respectively. The event that occurs may be a deterioration of a member of variant $i$, ($N_i\to N_i -1$), chosen with probability $N_i\omega_i/\Lambda$, or production of variant $j$ due to replication of variant $i$, ($N_j\to N_j+1$), chosen with probability $N_i\alpha_i\mu_{ij}/\Lambda$, such that the probabilities of all possible events sum to $1$.
\par
Application of such rules leads to a simulation of the system where $N(t)$ varies in time from its initial value $N(0)=N$. As such, in order to investigate the effects of a particular $N$, we then modify these dynamics such that they remain close to $N(t)\simeq N=N(0)$ throughout the simulation. This is achieved by dynamically modifying the replication rates, $\alpha_i\to \tilde{\alpha}_i(t)$, according to
\begin{align}
\tilde{\alpha}_i(t) = \alpha_i\left[\lambda - \frac{1}{1+e^{-\kappa(N(t)-N(0))}}\right].
\end{align}
Here $\kappa$ and $\lambda$ are then parameters, which we set as $\lambda=3/2$ and $\kappa=10/N(0)$, such that $\tilde{\alpha}_i(t)=\alpha_i$ when $N(t)=N(0)=N$. This will then be a fixed point of the system in cases where we have $\langle\alpha\rangle_{\mathbf{n}}\sim \langle\omega\rangle_{\mathbf{n}}$, which is the case in our examples.


\end{document}